\documentclass[fleqn,usenatbib]{mnras}

\usepackage[T1]{fontenc}

\DeclareRobustCommand{\VAN}[3]{#2}
\let\VANthebibliography\thebibliography
\def\thebibliography{\DeclareRobustCommand{\VAN}[3]{##3}\VANthebibliography}

\usepackage{graphicx}	
\usepackage{amsmath}	
\usepackage{extarrows} 
\usepackage{mathtools}

\title[Agnostic Model of Exoplanet Photosynthesis]{An \emph{Agnostic} Machine Learning Model of Photosynthetic Habitability}

\author[C. Gray et al]{
Callum Gray$^{1}$\thanks{E-mail: callum.gray@qmul.ac.uk}, 
Cassandra Hall$^{2}$, 
Stefano Santabarbara$^{3}$, 
Klaus Schmidt-Rohr$^{4}$, \newauthor
Andrew Ringham$^{5}$, 
Edward Gillen$^{5}$, 
Thomas J. Haworth$^{5}$ 
and Christopher D. P. Duffy$^{1}$\thanks{E-mail: c.duffy@qmul.ac.uk}
\\
$^{1}$ Digital Environment Research Institute (DERI), Queen Mary University of London, London E1 1HH, United Kingdom\\
$^{2}$ Department of Physics and Astronomy, Franklin College of Arts and Sciences, University of Georgia, United States of America\\
$^{3}$ Institute of Agricultural Biology and Biotechnology, National Research Council of Italy, Milan, Italy\\
$^{4}$ Department of Chemistry, Brandeis University, MS 015. 415 South Street Waltham, MA 02453-9110, United States of America\\
$^{5}$ Astronomy Unit, Queen Mary University of London, Mile End Road, London E1 4NS, United Kingdom
}

\date{Accepted XXX. Received YYY; in original form ZZZ}

\pubyear{\the\year{}}

\begin{document}
\label{firstpage}
\pagerange{\pageref{firstpage}--\pageref{lastpage}}
\maketitle

\begin{abstract}
The search for exoplanet biosignatures is guided by whether planetary environments can sustain photosynthesis. As such, the Photosynthetic Habitable Zone (PHZ) was recently proposed, as the overlap between the canonical habitable zone and the orbital range where stellar irradiance is sufficient to drive photosynthesis. Existing PHZ estimates rely on empirical light–response curves from Earth phytoplankton, and thus include implicit Earth-centric biases.

We introduce an \emph{agnostic} PHZ derived from a generalized model of photosynthesis grounded in thermodynamics and redox chemistry, without reference to model organisms. The model is built on a generic photochemical reaction in which photon capture couples oxidation of a donor molecule to the reduction of $CO_2$. The optical properties and $CO_2$ reduction rate are optimized against irradiance spectra for exoplanets orbiting main-sequence stars, using a genetic algorithm that mimics evolution by natural selection.
Our simulations predict that photosynthetic organisms compensate for reduced flux by evolving larger light-harvesting structures. As a result, photosynthetic viability declines only linearly with orbital distance, despite stellar flux falling off quadratically. As such, the agnostic PHZ expands well beyond previous Earth-based estimates. Earth-like (visible light) oxygenic photosynthesis is flux-limited at the outer habitable zone for cool M-dwarf stars; however, both anoxygenic photosynthesis and a hypothetical, NIR-driven oxygenic photosynthesis are viable across the entire habitable zone for M, K, and G stars. This implies that M-dwarf exoplanets could sustain robust oxygenic photosynthesis, though it would be different to that found on Earth, presenting reflectance biosignatures in the NIR band rather than the visible.
\end{abstract}

\begin{keywords}
Astrobiology --- Biosignatures --- Exoplanets --- M dwarf stars --- Xenobiology
\end{keywords}


   
\section{Introduction}

The Habitable Worlds Observatory (HWO) was identified as the top-priority flagship mission in the most recent astrophysics decadal survey, Astro2020. It will be the first telescope designed specifically to search for signs of life on other planets, capable of retrieving both atmospheric \citep{barbosa2025towards}  and surface evidence of oxygen \citep{borges2024detectability}. As such, it is of critical importance to determine a target list of candidate exoplanets that are capable of supporting oxygenic photosynthesis. 

\citet{hall2023new} recently defined the \emph{Photosynthetic Habitable Zone} (PHZ) as the sub-division of the habitable zone in which the incident photon flux can support oxygenic photosynthesis. The sufficiency of photon flux is based on the light curve for Earth phytoplankton, a function that maps light intensity onto the $CO_2$ fixation rate \citep{yang2020quantifying}. At low intensities $CO_2$ fixation increases with increasing flux, before saturating and then declining due to the onset of photo-damage\citep{stirbet2020photosynthesis}. 

Oxygenic photosynthesis on Earth requires photons in the 400--700~nm spectral window, termed \emph{photosynthetically active radiation} or PAR \citep{meek1984generalized}. The \emph{red limit} at 700~nm is enforced by the minimum energetic requirements of $H_2O$ oxidation, though there are some species of cyanobacteria \citep{nurnberg2018photochemistry} and plants \citep{hao2025utilising} that can extend this to 750~nm, albeit growing very slowly in only 700--750~nm light \citep{nien2022use}. Earth plants typically need between 10--300~$\mu$mol photons m$^{-2}$ s$^{-1}$ of PAR \citep{ritchie2010modelling}, though this varies a lot between species. Algae and cyanobacteria only need 1-10 $\mu$mol photons m$^{-2}$ s$^{-1}$ \citep{jodlowska2013combined} and recently \citet{hoppe2024photosynthetic} measured algal growth under the arctic icepack at fluxes of 0.04~$\mu$mol photons m$^{-2}$ s$^{-1}$, only slightly higher than the theoretical lower limit of 0.01~$\mu$mol photons m$^{-2}$ s$^{-1}$ \citep{raven2000put}. Upper limits are similarly variable, with some high light-adapted plants (often cacti) tolerating $>$ 1000~$\mu$mol photons m$^{-2}$ s$^{-1}$ \citep{10.1093/conphys/cox042}, while high-light stress can occur at as low as 200~$\mu$mol photons m$^{-2}$ s$^{-1}$ for some cyanobacteria \citep{islam2017growth}. 

\begin{figure*}
    \centering
    \includegraphics[width=\textwidth]{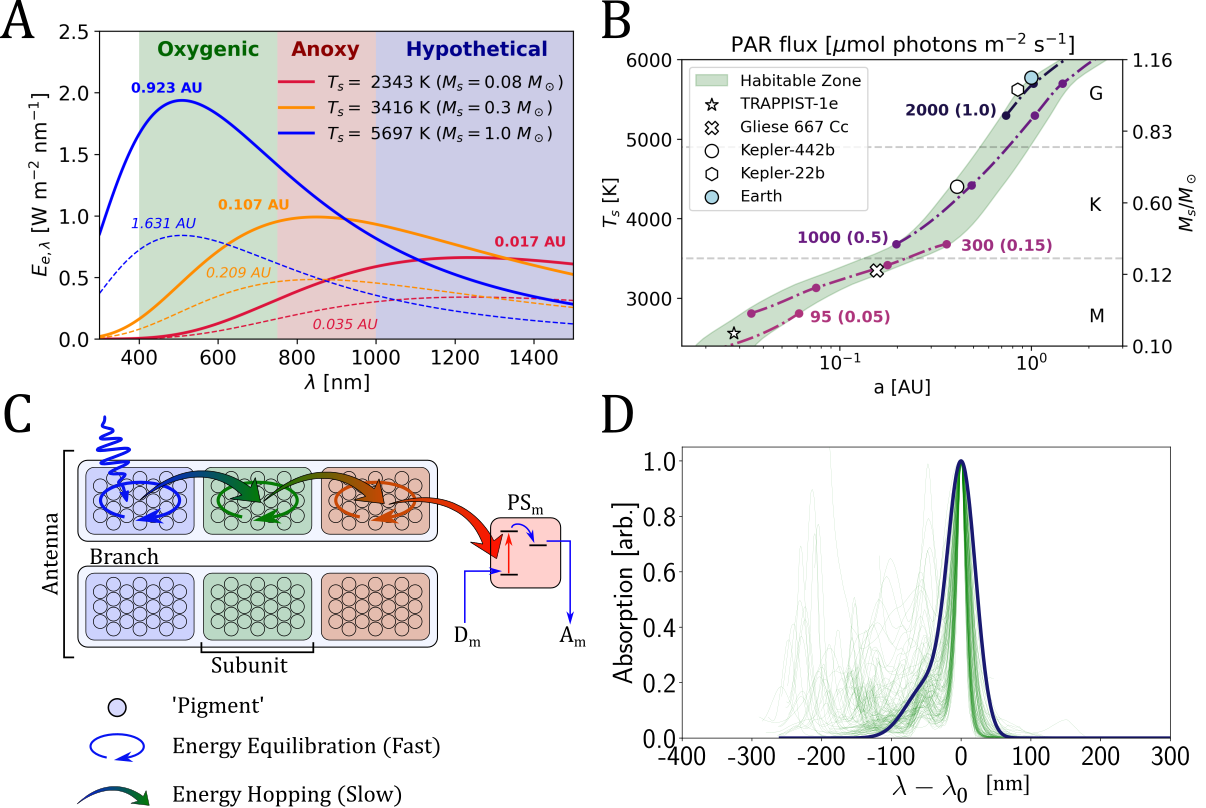}
    \caption{\textbf{A}: The model spectral irradiances, $E_{e,\lambda}$, for an exoplanet at the inner (solid lines) and outer (dashed) edges of the habitable zone around main sequence stars of different masses/temperatures ($M_{s}$ and $T_{s}$, respectively). As in \citet{hall2023new}, we assume that the stars are black body emitters and neglect attenuation by the planet atmosphere. The green shaded region indicates the 400--750~nm band used by oxygenic photoautotrophs on Earth, purple shading indicates the 800--1000~nm band used by Earth's anoxygenic photoautotrophs, and the blue shading indicates the NIR band that could be used to drive various \emph{hypothetical} forms of oxygenic photosynthesis \citep{wolstencroft2002photosynthesis}. \textbf{B}: The total average PAR flux (400--750~nm) incident on an exoplanet as a function of stellar effective temperature ($T_s$) (stellar mass, $M_s$, also shown) and the orbital semi-major axis $a$. The dashed lines show contours of constant PAR flux with the numbers showing the absolute flux in $\mu$mol photons m$^{-2}$ s$^{-1}$ (relative fluxes in brackets). The shaded green region indicates the habitable zone as defined in \citet{hall2023new} and for reference several worlds from the Habitable Worlds Catalogue \citep{HWCArecibo} are listed alongside Earth. \textbf{C}: A schematic of the \emph{antenna} model reported in \citet{chitnavis2024optimizing} and \citet{gray2025predicting}. \emph{Pigments} (circles) are generalized $\pi$-conjugated molecules defined entirely by their absorption lineshapes. A strongly-coupled cluster of identical pigments forms a \emph{subunit}, a chain of subunits forms a \emph{branch}, and the antenna can have several identical branches. Pigments absorb incident photons, with the resulting \emph{exciton} rapidly equilibrating over the local subunit (circular arrows). The exciton then \emph{hops} via resonant energy transfer (curved arrows) between subunits and eventually arrives at the \emph{reaction centre} (pink square) where it is used to drive the light reactions. \textbf{D}: A schematic of the absorption lineshape of our generalized `pigments' (dark blue), obtained by averaging over 116 organic pigment molecules (faint green) taken from the PhotochemCAD database~\citep{du1998photochemcad,taniguchi2020absorption}. The peak wavelength of the \emph{phononless} line, $\lambda_{0}$, is a free parameter in subsequent calculations.}
\label{fig:intro}
\end{figure*}

What does the diversity of PAR requirements/tolerances mean for astrobiology? Fig. \ref{fig:intro} \textbf{A} shows black-body approximations of the top-of atmosphere spectral irradiances, $E_{e,\lambda}$, for planets at the inner and outer edges of the habitable zones for various stars, calculated as in \citet{hall2023new}. The green `Oxygenic' area denotes the PAR region, and PAR flux decreases steeply with decreasing stellar temperature, $T_s$. Fig. \ref{fig:intro} \textbf{B} shows the habitable zone overlaid with contours of constant integrated PAR flux as a function of both $T_s$ and orbital semi-major axis, $a$. For the coolest M-dwarf stars we expect PAR fluxes of $\sim$ 100--300~$\mu$mol m$^{-2}$ s$^{-1}$, roughly 5--15\% of the equivalent maximum around a G2-type star. Of course these fluxes would be attenuated by atmospheric/environmental conditions, but it seems reasonable to assume that some form of oxygenic photosynthesis is feasible across the entire main sequence habitable zone. \citet{hall2023new}, however, presented a more sophisticated argument based on the model of \citet{eilers1988model}, in which \emph{photosynthetic rate} is expressed as an empirical function of incident PAR intensity, $I$,
\begin{align}\label{eq:PI}
P_\text{rate}(I) &= \frac{1}{\alpha I^2 + \beta I + \gamma} - R_\text{rate} 
\end{align}
where parameters $\alpha$, $\beta$ and $\gamma$ are taken from phytoplankton \citep{yang2020quantifying}. $R_\text{rate}$ is the \emph{dark respiration} rate, essentially the minimum metabolic demand of the organism; the PHZ is defined by $P_\text{rate}(I) > 0$. For anything less than excellent conditions (low atmospheric flux attenuation and small $R_\text{rate}$) Eqn.~(\ref{eq:PI}) predicts a PHZ significantly narrower than the habitable zone. Still, five planets were identified as robust candidates for supporting oxygenic photosynthesis: Kepler-452 b, Kepler-1638 b, Kepler-1544 b, Kepler-62 e, and Kepler-62 f. However, we should remember that these limits are derived from Earth organisms that evolved in a PAR-rich environment \citep{matsuo2025archaean} and probably don't reflect absolute physical constraints. We therefore need an \emph{agnostic} model of photosynthesis, one that is based on basic physical/chemical constraints rather than parameters (explicit and implicit) taken from model organisms.

Often neglected in astrobiological models of photosynthesis is the \emph{light-harvesting antenna}, an evolutionary strategy adopted by all of Earth's photosynthetic organisms which enhances their capacity to capture light. Fig. \ref{fig:intro} \textbf{C} sketches an abstract schematic of an antenna, essentially an assembly of different pigment-binding proteins, additional to and associated with the \emph{reaction centre} (RC) complexes that carry out the photosynthetic \emph{light reactions}. Antenna structure is extremely diverse. For example, the antenna of plants is a constantly-rearranging network of chlorophyll-binding proteins, switching between light-harvesting and photoprotective `modes' in low and high light respectively \citep{johnson2011photoprotective}. Conversely, cyanobacteria have a larger, less flexible antenna which binds a greater variety of pigments and appears to be tuned to maximizing light capture in low PAR flux \citep{glazer1985light}. Regardless, all antennae operate on the same principles: They bind various pigments at high concentration, arranged in an \emph{energy funnel} where pigments are progressively blue-shifted as they get further away from the RC \citep{lokstein2021photosynthetic}. Photon absorption results in a mobile excitation (an `exciton') that hops between neighbouring pigments (via \emph{resonant energy transfer}; \citealt{scholes2006energy}) and diffuses down the funnel to the RC. In \citet{chitnavis2024optimizing} we developed a generalized antenna model and showed that, while there is a upper limit to antenna size (above which they stop working), oxygenic organisms evolving around M-dwarfs could \emph{partially} compensate for reduced PAR flux by evolving larger, more complex antennae. In \citet{gray2025predicting} we coupled this antenna model to an evolutionary machine learning algorithm which optimizes antenna structure to different fluxes via a process that mimics natural selection. By exposing this `learning antenna' to different light environments on Earth we were able to reproduce the various antenna structures seen in nature. Applied to main sequence stars it predicts that some form of oxygenic photosynthesis could evolve even around cool M-dwarfs, though it may require a very large antenna and would reach only half the efficiency seen on Earth \citep{chitnavis2024optimizing}.              

We should also consider entirely different forms of photosynthesis. Even on Earth, organisms like purple \citep{hunter2009purple} and green sulphur bacteria \citep{frigaard2008genomic} carry out \emph{anoxygenic} photosynthesis, a process which uses Near Infrared (NIR) light to oxidize geochemical reductants rather than $H_2O$. Though anoxygenic photosynthesis currently contributes less than 1\% of the global carbon fixation \citep{johnston2009anoxygenic}, it dominated on Archean Earth \citep{olson2006photosynthesis}, and remains important in arid or PAR-limited environments \citet{bay2021chemosynthetic}. Various models have predicted that the NIR-dominant flux around an M-dwarf would likely select for biospheres based on robust anoxygenic photosynthesis, rather than struggling oxygenic organisms \citep{lehmer2021peak,duffy2023photosynthesis}. Importantly, these biospheres could still be detected despite the lack of an $O_2$ atmosphere. \citet{10.1093/mnras/stae601} have shown that analogues of Earth's purple bacteria could yield a surface reflectance \emph{edge} in the NIR band, most recently extending this to cloud based populations \citep{coelho2025colors}.

Finally, it is also possible that photosynthetic processes with no Earth analogue could evolve given the right conditions.  \citet{bains2014photosynthesis} proposed a hypothetical, NIR-driven \emph{hydrogenic} ($CH_4$ oxidising) photosynthesis that could dominate on Hycean worlds \citep{2021ApJ...918....1M}. Similarly, \citet{wolstencroft2002photosynthesis} proposed a form of oxygenic photosynthesis that can drive $H_2O$ oxidation with NIR light, based on introducing intermediate steps that can combine the energy of several photons. 

Here, we attempt to bring all of these ideas together to define an \emph{agnostic} PHZ. We present a general thermodynamic model that treats oxygenic, anoxygenic and even hypothetical forms of photosynthetic light reactions in a consistent manner. We then combine this with the `learning antenna' model benchmarked in \citep{chitnavis2024optimizing,gray2025predicting}, and predict how different forms of photosynthesis perform across the habitable zone when we are free of any parameter constraints derived from Earth organisms.     

\section{Methods}

\subsection{Defining the habitable zone and spectral irradiances}

Our models learn to maximize \emph{photosynthetic fitness} in an environment defined by the average spectral irradiance $E_{e,\lambda}(a,T_{*}, R_{*})$ (in units of W $m^{-2}$ nm$^{-1}$) at the surface of a planet orbiting distance $a$ from a star of effective temperature $T_*$ and radius $ R_*$.

We begin as in~\citet{hall2023new} by using the stellar isochrones of~\citet{baraffeNewEvolutionaryModels2015} to obtain stellar effective temperatures, radii and luminosities for stars of different masses at age 4Gyr, then use the method of~\citet{kopparapuHABITABLEZONESMAINSEQUENCE2013} to calculate the limits of the HZ for each chosen star. Finally we sample at intervals between the inner and outer limits of the HZ and calculate the irradiance as a function of wavelength $\lambda$ received by a planet at a given radius $ a $ from a star with mass and radius $ T_{*}, R_{*} $ by,
\begin{equation}
    E_{e, \lambda}(a, T_{*}, R_{*}) = \frac{2\pi h c^2}{\lambda^5} \left(\frac{R_{*}}{a}\right)^2 \left[ \exp{\frac{hc}{\lambda k_B T_{*}}} - 1\right]^{-1} 
\end{equation}

where $h = 6.626\times10^{-34} \text{Js}$ is Planck's constant, $c = 3\times10^8$ms$^{-1}$ is the speed of light in vacuum and $k_B = 1.38\times10^{-23} \text{J K}^{-1}$ is Boltzmann's constant.

\subsection{Generalized photosynthetic antenna systems}

The antenna component of our model is outlined in \citet{chitnavis2024optimizing} and briefly summarized here. Antennae are composed of `\emph{pigments}' which are assumed to be $\pi$-conjugated organic molecules. In our model, they are solely defined by their wavelength-dependent absorption spectrum $A(\lambda$), for which we define a generalized \emph{line-shape} function. The red-most absorption band of organic pigments typically features an intense Gaussian peak (the \emph{vibrationless} line) plus a series of progressively blue-shifted and progressively weaker \emph{vibronic shoulders} (see Fig. \ref{fig:intro} \textbf{D}). We approximate this with a two-peak line-shape function,
\begin{align}\label{eq:lineshape}
\begin{split}
A(\lambda) &=  B_0 \exp\left(-\frac{(\lambda-\lambda_{0})^2}{2{w_0}^2}\right)\\
&+ \left(B_0-\Delta B\right)\exp\left(-\frac{(\lambda-\lambda_{0}-\Delta \lambda)^2}{2\left(w_0\pm \Delta w\right)^2}\right)
\end{split}
\end{align}
The first term is the vibrationless peak, centered on $\lambda_0$ with width $w_0$ and amplitude $B_0$, and the second term is the vibronic shoulder. To parameterize $A(\lambda)$ we sampled all `Natural Chlorophylls' and `Naturally Derived Porphyrins' in the \emph{PhotochemCAD} pigment database \citep{du1998photochemcad,taniguchi2020absorption}. Fig. \ref{fig:intro} \textbf{D} plots these sample spectra, shifted so that they have a common $\lambda_0$ and scaled so that all $B_0 = 1$, as faint green lines. $w_0$, $\Delta B$, $\Delta \lambda$ and $\delta w$ are determined by averaging over the sample, producing the $A(\lambda)$ plotted as a thick blue line. This is then normalized so that $\int_0^\infty d\lambda A(\lambda)=1$ and $\lambda_0$ is treated hereafter as a free parameter. 

Clusters of pigments with the same $\lambda_0$ form `\emph{subunits}', multiple subunits can be arranged into a linear `\emph{branch}', and the antenna can be composed of several identical branches. Fig. \ref{fig:intro} \text{C} shows an antenna composed of two branches each containing three subunits with different $\lambda_0$ (denoted by different colours). However, the number of pigments and their $\lambda_0$ in each subunit, the number of subunits per branch and the number of branches are all free parameters to be optimized by the learning model.    

The basic sequence of events in light-harvesting is as follows: One of the antenna pigments absorbs a photon and the resulting exciton is transferred between pigments. We assume a hierarchy of scale in which exciton delocalization within a subunit (circular arrows in Fig. \ref{fig:intro} \textbf{C}) is much faster than hopping between subunits (arched arrows). The rate of photon absorption by antenna subunit $s$ is given by
\begin{align}
    \gamma_{s}=\sigma_{s,p}N_{s,p}\int_0^\infty d\lambda \frac{\lambda}{hc} E_{e,\lambda} A\left(\lambda; \lambda_{0,s}\right)
\end{align}
where $\sigma_{s,p}$ is the \emph{integrated absorption cross-section} of a single pigment (typically $\sigma \sim 10^{-20}$ cm$^{-2}$ for organic pigments \citep{Noy_Chla_sigma}) and $N_{s,p}$ is the number of pigments in the subunit. We use a normalised absorption line shape together with a per-pigment integrated cross section to facilitate working with antennae that have variable numbers of pigments and absorption peaks. The rate constants for exciton hopping between subunits $s$ and $s'$ are related by the \emph{detailed balance} condition,
\begin{align}\label{eq:detailed}
\frac{k_{s'\rightarrow s}}{k_{s\rightarrow s'}}=\exp\left(\frac{1}{k_BT} \left(\Delta H_{s\rightarrow s'}-T\Delta S_{s\rightarrow s'}\right)\right)   
\end{align}
where, 
\begin{align}
\Delta H_{s\rightarrow s'} = hc\left(\frac{1}{\lambda_{0,s'}}-\frac{1}{\lambda_{0,s}}\right)    
\end{align}
is the enthalpy change and,
\begin{align}
\Delta S_{s\rightarrow s'} = k_B \ln \left(\frac{N_{s'}}{N_{s}}\right)    
\end{align}
Eqn. (\ref{eq:detailed}) enforces a thermodynamic direction on energy transfer: Energy transfer from blue-shifted to red-shifted subunits is favoured, as is transfer from a smaller subunit to a larger one. The overall process of harvesting light intrinsically \emph{reduces} entropy, as it involves concentrating energy from a large antenna into a small RC. However, the energy funnel structure ensures overcompensation by an even larger reduction in enthalpy. It is the intrinsic entropy reduction that prevents arbitrarily large antenna structures from working effectively \citep{chitnavis2024optimizing}.

\subsection{Generalized model of photosynthetic light reactions}

\begin{figure*}[htp!]
    \centering
    {\includegraphics[width=0.75\textwidth]{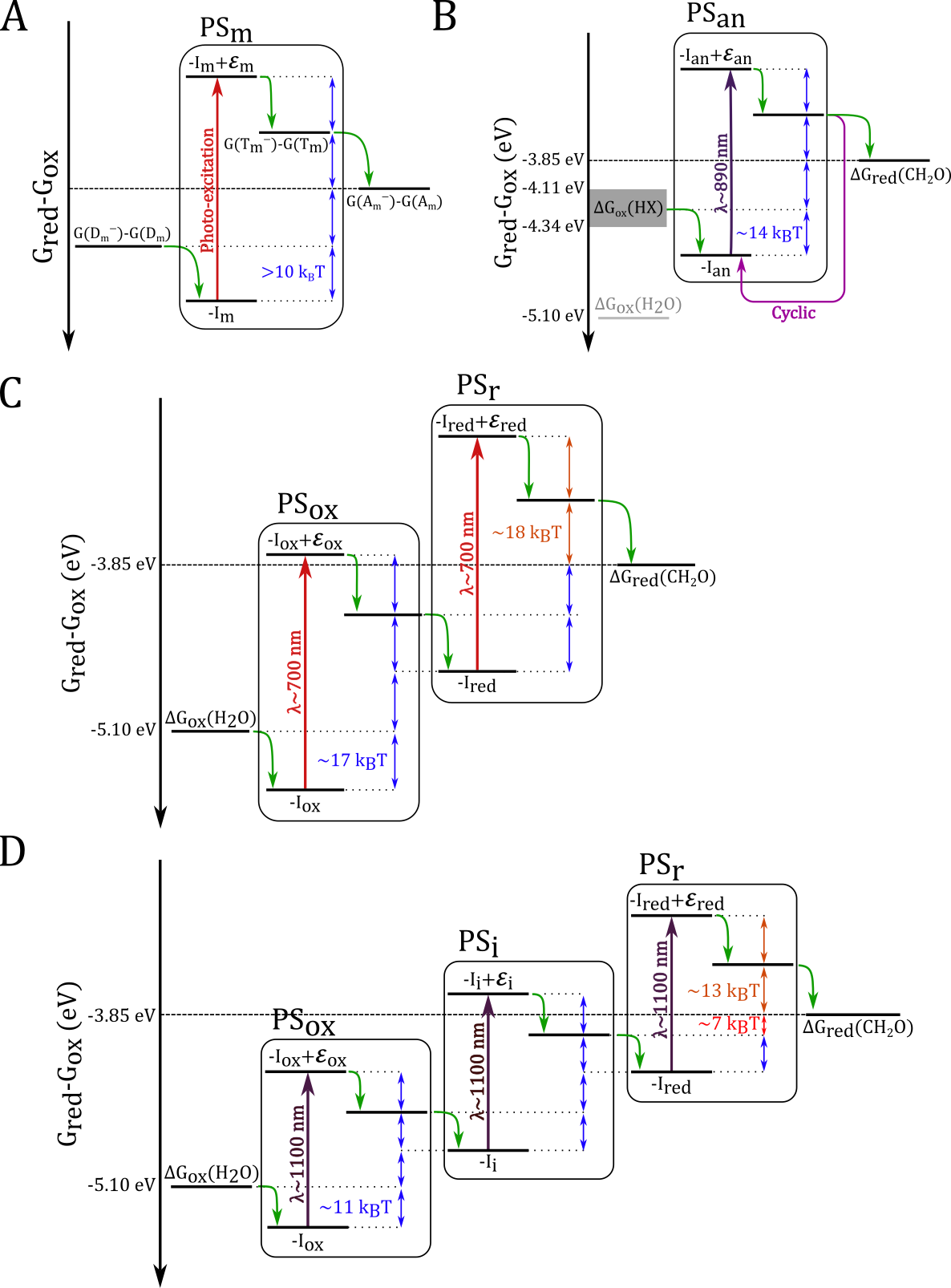}}
     \caption{\textbf{A}: An energy level diagram of the generalized \emph{photosystem} $PS_m$. The levels indicate one-electron reduction potentials for various half-reactions. The reaction centre pigment undergoes photo-excitation followed by charge separation via electron transfer to a \emph{trap} molecule $T_m$. The oxidized pigment then oxidizes a donor $D_m$ and the reduced trap, $T_m^{-}$, reduces an electron acceptor $A_m$. The faded gray lines indicate that there may be several intermediate charge separation steps so $T_m$ should be interpreted as an effective state. The four energy gaps shown should be $> 10 k_BT$ to ensure forward redox processes are irreversible and that $D_m$ cannot reduce $A_m$ directly. \textbf{B}: A schematic of an \emph{anoxygenic} supersystem which oxidizes some geochemical reductant $HX$ and reduces $CO_2$. Since $HX$ is assumed to be more readily oxidized than $H_2O$ (grey line) a single photosystem, $PS_{an}$, using light of $\lambda \leq 890$ nm is sufficient. Cyclic electron flow is also possible (purple line) which generates \emph{chemical energy} but does not reduce $CO_2$. For consistency with the other models, we assume that two photons are needed for one-electron reduction of $CO_2$: one to generate a unit of chemical energy via cyclic flow and one to move an electron to from $HX$ t $CO_2$. \textbf{C}: An Earth-like \emph{oxygenic} supersystem that uses red light ($\lambda \leq 700$~nm), requiring two photosystems connected in series: one oxidising, $PS_{ox}$, and one reducing, $PS_{r}$. Transfer of an electron from $PS_{ox}$ to $PS_r$ generates chemical energy which we assume contributes to the one-electron reduction of $CO_2$. \textbf{D}: A \emph{hypothetical} oxygenic supersystem that can oxidize $H_2O$ and reduce $CO_2$ with NIR light ($\lambda \leq 1100$ nm) via three photosystems, $PS_{ox}$, $PS_i$, and $PS_r$, connected in series.}
      \label{fig:levels}
\end{figure*}

Energy delivered to the RCs is used to drive the coupled oxidation of a donor and reduction of an acceptor. We develop an agnostic description of this by generalizing the \emph{extended Z-scheme} of oxygenic photosynthesis developed by \citet{schmidt2021}. We define a general \emph{photosystem} $PS_m$ as shown in Fig. \ref{fig:levels} \textbf{A} which consists of a chain of redox processes:
\begin{enumerate}
\item{A special low-energy pigment is \emph{photo-excited} via the antenna: $P_m \rightarrow P_m^*$}
\item{The excited pigment undergoes \emph{charge separation} via electron transfer to an electron \emph{trap}: $P_m^*+T_m\rightarrow P_m^+ + T_m^-$}
\item{The reduced trap then reduces the electron acceptor: $T_m^-+A_m \rightarrow T_m + A_m^-$.}
\item{The oxidized pigment then oxidizes the electron donor, $P_m^++D_m\rightarrow P_m + D_m^+$}
\end{enumerate}
Note that Steps 3. and 4. don't have a strict time-ordering.

Though we specify no molecular detail, there are several thermodynamic constraint on such a photosystem. Firstly, excitation of $P_m$ results in a \emph{free energy} increase,
\begin{align}
    G(P_m^*) - G(P_m) = \varepsilon_m\left(1-\frac{4T_p}{3T_\star}\right)
\end{align}
where $\varepsilon_m=hc/\lambda_{0,m}$ is the photon energy and $T_p$ is the average temperature of the planet surface. As shown by \citet{schmidt2021} this includes a contribution from the entropy decrease associated with the annihilation of a photon \citep{delgado2017entropy},
\begin{align}\label{eq:Carnot}
 \Delta S_\gamma = \frac{4}{3T_\star}\left(\frac{hc}{\lambda_{0,m}}\right)   
\end{align}
This is often neglected since $T_p/T_\star$ is typically small, though not vanishingly so for lower temperature/mass stars. For consistency with  \citet{chitnavis2024optimizing} we make the approximation,
\begin{align}
    G(P_m^*) - G(P_m) \sim \varepsilon_m
\end{align}

The second constraint is that charge separation will only occur if it lowers free energy,
\begin{align}\label{eq:exergonic}
\Big[G(P_m^+)+G(T_m^-)\Big]-\Big[G(P_m^*)+G(T_m)\Big] \leq 0    
\end{align}
It is more intuitive to rearrange Eqn. (\ref{eq:exergonic}) as,
\begin{align}
\Big[G(T_m^-)-G(T_m)\Big]-\Big[-I_m + \varepsilon_m\Big] \leq 0
\end{align}
where $I_m = G(P_m^+)-G(P_m)$ is the ionization potential of $P_m$. 

Similarly, donor oxidation and acceptor reduction are only spontaneous if,
\begin{align}
-I_m-\Big[G(D_m^-) - G(D_m)\Big] &< 0
\end{align}
and,
\begin{align}
\Big[G(A_m^-) - G(A_m)\Big]-\Big[G(T_m^-) - G(T_m)\Big] &< 0    
\end{align}
respectively. 

Finally, we assume,
\begin{align}
\Big[G(A_m^-) - G(A_m)\Big]-\Big[G(D_m^-) - G(D_m)\Big] > 0    
\end{align}
In other words, the donor should not be able to spontaneously reduce the acceptor, since that would bypass any need for photo-excitation (it would be spontaneous \emph{chemosynthesis}). 

The energy differences between the reduced and oxidized species, $\left[G(X_m^-)-G(X_m)\right]$, are one-electron reduction potentials and they constitute the energy levels of $PS_m$. We assume the sub-processes in $PS_m$ should not only be spontaneous, but thermodynamically irreversible, so that they out-compete waste processes that defeat the point of the photosystem. Charge separation has to out-compete excited state decay, 
\begin{align}
P_m^* \rightarrow P_m + \text{heat}   
\end{align} 
and acceptor reduction has to out-compete \emph{non-radiative charge recombination}  
\begin{align}
P_m^+ + T_m^- \rightarrow P_m + T_m+\text{heat}   
\end{align} 
Therefore, the associate energy gaps should be thermally large,
\begin{align}
\Big[G(P_m^+)+G(T_m^-)\Big]-\Big[G(P_m^*)+G(T_m)\Big] \leq -10 k_BT        
\end{align}
\begin{align}
\Big[G(A_m^-) - G(A_m)\Big]-\Big[G(T_m^-) - G(T_m)\Big] &< -10 k_BT        
\end{align}
and so on. This energetic scheme is shown in Fig. \ref{fig:levels} \textbf{A}

This framework can be extended to chains of multiple photosystems (hereafter called photosynthetic \emph{supersystems}), and we model a generalized anoxygenic photosynthesis, an Earth-like oxygenic photosynthesis, and a hypothetical form of NIR-driven oxygenic photosynthesis inspired by \citet{wolstencroft2002photosynthesis}. 

Fig \ref{fig:levels} \textbf{C} shows an Earth-like \textbf{oxygenic supersystem} which uses red light (700~nm). The ultimate electron donor is $H_2O$ and while real $H_2O$ oxidation requires a multi-electron catalytic cycle, we can define the effective \emph{one-electron} half-reaction,
\begin{align}
\frac{1}{2}H_2O \rightarrow \frac{1}{4}O_2 + H^+(\text{aq}) + e^-
\end{align}
with a one-electron reduction potential \citep{schmidt2021},
\begin{align}
\begin{split}
\Delta G_{ox}(H_2O)&= G\left(\frac{1}{2}H_2O\right)-G\left(\frac{1}{4}O_2+H^+\right)\\ 
&\sim - 5.1 \text{ eV}   
\end{split}
\end{align}
The ultimate electron acceptor is $CO_2$, which in real systems generally also requires complex catalysis (e.g. the Calvin-Benson cycle, as reviewed in \citet{gurrieri2021calvin}). Again, we define an effective one-electron half-reaction,
\begin{align}
\frac{1}{4}CO_2 + H^+(\text{aq}) \rightarrow \frac{1}{4}CH_2O + \frac{1}{4}H_2O
\end{align}
with a reduction potential \citep{schmidt2021},
\begin{align}
\begin{split}    
\Delta G_{red}(CH_2O)&=G\left(\frac{1}{4}CH_2O+\frac{1}{4}H_2O\right)\\
&-G\left(\frac{1}{4}CO_2+H^+\right)\\
& \sim -3.85 \text{ eV}
\end{split}
\end{align}
From the Fig. \ref{fig:levels} \textbf{C} we see that it is not possible for a single red light photosystem to both oxidize $H_2O$ and reduce $CO_2$. While, 
\begin{align}
\frac{hc}{700\text{ nm}} > \Delta G_{red}(CH_2O) - \Delta G_{ox}(H_2O)    
\end{align}
a sizable fraction of the photon energy needs to be dissipated to ensure that charge separation, $H_2O$ oxidation, etc. are irreversible. Two photosystems with different ionization potentials, connected in series, are needed: a $H_2O$-oxidizing system, $PS_{ox}$, and $CO_2$-reducing system, $PS_{red}$. This is one of the arguments presented by \citet{schmidt2021} for why oxygenic photosynthesis requires two chemically distinct photosystems: The water-oxidizing Photosystem II (PSII) and the reducing Photosystem I (PSI). 

The \textbf{anoxygenic supersystem} is shown in Fig. \ref{fig:levels} \text{B}. The donor is assumed to be some species $HX$ which is more easily oxidized than $H_2O$ and therefore a single photosystem, $PS_{an}$, absorbing longer wavelength light, is sufficient. Anoxygenic photosynthesis on Earth is rather diverse, both in terms of the donor and RC structure \citep{li2024review}. However, they all use photons in the 800--1000~nm range and the model in Fig. \ref{fig:levels} \textbf{B} is a generalization, with $\Delta G_{ox}(HX) \sim -4.34 - -4.11 $ eV and $\varepsilon_{an} \sim hc/(890 \text{ nm})$. 

The \textbf{hypothetical supersystem} is shown in Fig. \ref{fig:levels} \text{D}. It is, in principle, possible to use NIR light ($\lambda \sim 1100$ nm) to oxidize $H_20$ with \emph{three} photosystems connected in series: $PS_{ox}$, $PS_{red}$ and an \emph{intermediate} system $PS_i$. 

In oxygenic photosynthesis on Earth, electron transfer between PSII and PSI drives the pumping of protons across a membrane. The resulting electrochemical gradient is then used to generate chemical energy in the form of ATP, which is required for the reduction of $CO_2$ via the Calvin-Benson cycle. In some anoxygenic organisms (such as purple bacteria) this same process is achieved via \emph{cyclic} electron flow about the single photosystem \citep{hu2002photosynthetic}. Oxygenic organisms also perform cyclic electron flow about PSI to generate additional chemical energy, particularly is stress conditions \citep{nawrocki2019mechanism}. We assume that our anoxygenic supersystem has to perform cyclic electron flow (purple arrow in Fig. \ref{fig:intro} \textbf{B}) for every alternate photon absorbed. One photon to provide the chemical energy via cyclic flow, and the other to provide the reducing power (the electron). We neglect cyclic electron flow in the oxygenic and hypothetical supersystems since linear flow produces both energy and reducing power. The aim is to treat these three models on an equal footing, and the role of cyclic flow in oxygenic photosynthesis will be addressed in future work.

Finally, while the models are as agnostic as possible, there are a number of parameters which set the basic timescale for processes. Rather than take these from specific organisms, we can make some general physical/chemical arguments. Firstly, excitation decay in organic molecules occurs on a timescale of $\sim 1-10$~ns \citep{lakowicz2006principles}, which means there would be a strong selection pressure for charge separation to be much faster. We therefore assume a charge separation timescale of $\sim 1-10$~ps. For the same reason, exciton transfer must occur on a timescale of $\sim 1-10$~ps, which is the case for molecular excitonic systems in general \citep{scholes2006excitons}. This base timescale is then tuned by the antenna structure via the detailed balance condition in  Eqn. (\ref{eq:detailed}). Redox processes, such as donor oxidation, $CO_2$ reduction and electron transfer between photosystems, are chemical reactions (involving intermediate carriers) and so are likely slower, typically $\sim 100\mu$s--10ms. Finally, charge recombination must be extremely slow (100ms--1s) and is therefore omitted from our model. Any photosynthetic system would face an extreme selection pressure to minimize the recombination rate since it completely undermines the purpose of the light reactions. A discussion of these timescales is given in Section 1.4 of the Supplementary Material. 

\subsection{Kinetic equations and the $CO_2$ reduction rate}

Having defined the antenna-plus-supersystem models we develop equations of motion (EOM) that are solved to predict the \emph{photosynthetic rate}, $\nu$. The EOM are a generalization of the antenna-only model presented in \citet{chitnavis2024optimizing} and \citet{gray2025predicting}, and are derived in full in the Section 1 of the Supplementary Material. In essence, we define a vector of time-dependent probabilities, associated with the different instantaneous configurations of the supersystem-plus-antenna. For example, the probabilities for the oxygenic supersystem-plus-antenna are,
\begin{align}
P(t) &\equiv P_{\mathbf{n}_x,\mathbf{n}^{ox},\mathbf{n}^{red}}(t)   
\end{align}
$\mathbf{n}_x$ is a vector of exciton occupancies, 
\begin{align}
\mathbf{n}_x=(n_1,n_2,\ldots,n_i,\ldots,n_N)    
\end{align}
where  $n_s$ denotes the number of excitons in the $s^\text{th}$ antenna subunit. $\mathbf{n}^{ox}$ is a set of indices denoting the state of $PS_{ox}$,
\begin{align}
 \mathbf{n}^{ox}=(n_{x}^{ox},n_{cs}^{ox},n_{-}^{ox},n_{+}^{ox})   
\end{align}
where
\begin{align}
\mathbf{n}^{ox}&=(0,0,0,0) = P_{ox}+T_{ox}\\  
\mathbf{n}^{ox}&=(1,0,0,0)=  P_{ox}^*+T_{ox}\\  
 \mathbf{n}^{ox}&=(0,1,0,0)=  P_{ox}^+ + T_{ox}^-\\  
\mathbf{n}^{ox}&=(0,0,1,0)=P_{ox}+T_{ox}^-\\    
 \mathbf{n}^{ox}&=(0,0,0,1)=P_{ox}^+ +T_{ox}
\end{align}
Similarly, $\mathbf{n}^{red}$ denotes the state of $PS_{red}$. 

The EOM contain terms for all possible processes that move the model between these different configurations,
\begin{align}\label{eq:dPdt}
\begin{split}    
\frac{d}{dt}P(t)&=\left[\frac{d}{dt}P(t)\right]_{\gamma}+\left[\frac{d}{dt}P(t)\right]_{ET}+\left[\frac{d}{dt}P(t)\right]_{D}\\
&+\left[\frac{d}{dt}P(t)\right]_{CS}+\left[\frac{d}{dt}P(t)\right]_{e^-}\\
&+\left[\frac{d}{dt}P(t)\right]_{H_2O}+\left[\frac{d}{dt}P(t)\right]_{CO_2}
\end{split}
\end{align}
The first line in Eqn. (\ref{eq:dPdt}) are the light-harvesting processes, with the $\gamma$, $ET$, and $D$ terms describing photon absorption, exciton transfer, and exciton decay respectively. The second line characterizes electron transfer, with $CS$ and $e^-$ terms describing charge separation and $PS_{ox}\rightarrow PS_{red}$ electron transfer respectively. Finally, the $H_2O$ and $CO_2$ terms characterize the $H_2O$ oxidation and $CO_2$ reduction steps. This formalism can be applied to the anoxygenic and hypothetical models with minor modifications to the $e^-$ term based on the number of photosystems. 

The set of EOM are solved in the steady state, 
\begin{align}
 \frac{d}{dt}P^{eq} = 0   
\end{align}
with a non-negative least-squares solver to give the equilibrium probabilities $P^{eq}$. From these we can calculate various observables such as the one-electron $CO_2$ reduction rate, $\nu$. 
 
\subsection{The genetic algorithm}

As in \citet{gray2025predicting} we optimize the antenna with a bespoke genetic algorithm. The full list of antenna parameters (number of branches, number of subunits per branch, number of pigments in each subunit, absorption peak $\lambda_0$ of each pigment) form a \emph{genome}, and the spectral irradiance, $E_{e\lambda}(a,T_s)$, constitutes the \emph{environment}. For a given $E_{e\lambda}(a,T_s)$ we select one of the photosynthesis models (denoted $\mathcal{M}$) and then define a population of $n_g$ random genomes,
\begin{align}
\mathbf{G}=\left(G_1,G_2,\ldots,G_m,\ldots,G_{n_g}\right)    
\end{align}
Using smaller $n_g$ will generally lead to faster simulations but a decrease in accuracy; larger $n_g$ will reduce the number of generations needed for convergence of the algorithm, but at the expense of computational complexity~\citep{lobo2007adaptive}. We set $n_g$ to 500 in this work, finding that this provided reasonable convergence times in most cases.

The algorithm then evaluates the photosynthetic \emph{fitness} of each genome,
\begin{align}
f(\mathbf{G},E_{e,\lambda},\mathcal{M}) = \nu(\mathbf{G},E_{e,\lambda},\mathcal{M}) -\chi N_p^a(\mathbf{G},E_{e,\lambda},\mathcal{M})
\end{align}
where $\nu$ is the $CO_2$ reduction rate, $\chi$ is the \emph{cost parameter} and $N_{p}^a$ is the total number of antenna pigments. $\chi$ is essentially the number of electrons per second, per pigment that have to be reinvested to maintain the photosynthetic apparatus. Therefore, $f(\mathbf{G},E_{e,\lambda},\mathcal{M})$  represents a biological imperative to do as much photosynthesis as possible with as small an antenna as possible, with $\chi$ setting the balance. \citet{gray2025predicting} showed that a value of $\chi \sim 0.02$ can reproduce the various antenna structures found on Earth, though the genetic algorithm is not very sensitive to small variations.

The different genomes are then stochastically selected according to their survival probability, 
\begin{align}
    p_m(\mathbf{G},E_{e,\lambda},\mathcal{M}) = \frac{\left(1-\exp\left(-\frac{F_m(\mathbf{G},E_{e,\lambda},\mathcal{M})}{F_{max}}\right)\right)}{\sum_m F_m}
\end{align}
where $F_{max}$ is the highest fitness across the population. The survivors rebuild the population via \emph{genetic crossover}, a process that randomly mixes the genome of two parents,
\begin{align}
 \mathbf{G}_m + \mathbf{G_n} \rightarrow \mathbf{G}_m + \mathbf{G_n} +\mathbf{G}_l    
\end{align}
using a procedure known as intermediate recombination~\citep{Eiben2015}, which works for real-valued genes by sampling from a uniform distribution centred around the parental values plus some random variation.

Finally, random mutations are applied to a randomly selected subset of the genomes,
\begin{align}
\mathbf{G}_m \rightarrow \mathbf{G'}_m    
\end{align}
for example changing the number of pigments in a subunit $N_{s, p}$, the absorption maximum of a subunit $\lambda_0$, or the number of branches in the antenna.

The process is then iterated until the average fitness of the population converges. We perform three repeats of our simulation for each $E_{e,\lambda}$ with different random seeds in order to average over different evolutionary histories; more repeats allow the system to probe different regions of the parameter space at the cost of simulation time.
\section{Results}
\begin{figure*}[ht!]
    \centering
    {\includegraphics[width=\textwidth]{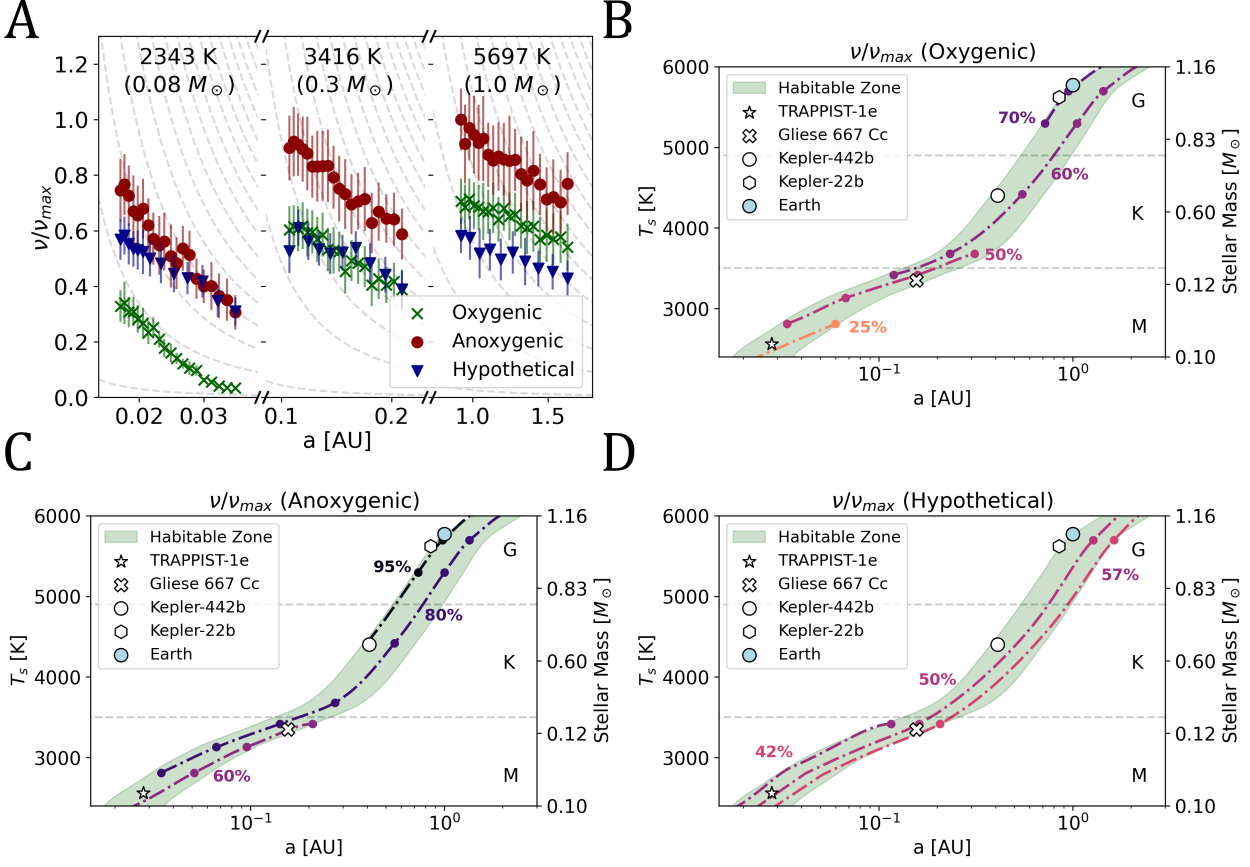}}
     \caption{\textbf{A}: The average \emph{relative} photosynthetic output rate, $\nu/\nu_\text{max}$, as a function of main sequence stellar temperature/mass and orbital semi-major axis, $a$, for oxygenic, anoxygenic and hypothetical forms of photosynthesis (denoted by green crosses, red circles and blue triangles respectively). The error bars represent variance in antenna structure in the final \emph{evolved} photosynthetic population output by the genetic algorithm. The faint dashed lines are a visual reference showing a $1/a^2$ dependence. \textbf{B}: The habitable zone (green) overlaid with contours of constant $\nu/\nu_\text{max}$ (coloured dashed lines) for \emph{oxygenic} photosynthesis. The contours are interpolated from the data points which are shown as filled circles and the corresponding values of $\nu/\nu_\text{max}$ are shown as a percentage. For reference, several candidates worlds from the Habitable Worlds Catalogue \citep{HWCArecibo} are listed, alongside Earth. \textbf{C}: Same as B but for anoxygenic photosynthesis. \textbf{D}: Same as B but for the \emph{hypothetical} NIR-driven oxygenic photosynthesis. Since only three stellar temperatures/masses were considered for the hypothetical form of photosynthesis (due to calculation cost), we added interpolated data points to generate the contours.}
      \label{fig:contours}
\end{figure*}

\subsection{Relative photosynthetic output}

In Fig. \ref{fig:contours} \textbf{A} we plot the average $CO_2$ rate (plus standard deviations as error bars) for the evolved populations of our three photosynthesis models across the habitable zone of different main sequence stars. We consider a range of stellar temperatures, $T_s$, across the main sequence but only tabulate three for clarity. We plot the \emph{relative} rate, $\nu/\nu_{max}$, where $\nu_{max}\sim 70$ electrons s$^{-1}$ is the highest rate across the entire data set. In terrestrial organisms the maximum possible rate is $\approx 100 \text{e}^{-} \text{s}^{-1} $, which is approximately the rate of the quinone reduction cycle in oxygenic PSII. As the orbital semi-major axis, $a$, increases $\nu/\nu_{max}$ declines \emph{linearly} while flux decreases as $\sim 1/a^2$ (indicated by faint dashed line). This reflects that the genetic algorithm tends to select for larger antennae in lower flux environments, as shown in Fig.~\ref{fig:costs}. However, antennae get less efficient as they increase in size, meaning this can only partially compensate for decreased flux.

The anoxygenic system performs very well across the entire range of main sequence stars, with $\nu/\nu_{max} < 0.5$ only for the outer part of the habitable zone around M-dwarf stars. Interestingly, the oxygenic model under-performs relative to the anoxygenic model even for the G-type stars. This is due to kinetic bottleneck of having two photosystems connected in series, a bottleneck that the anoxygenic system lacks. Around M-dwarf stars, the oxygenic system is strongly light-limited, and becomes barely viable towards the outer edge of the habitable zone. Finally, the hypothetical system suffers from the significant kinetic bottleneck of having \emph{three} photosystems connected in series, delivering at most $\nu/\nu_{max}\sim 0.6$. It performs worse that the oxygenic model around the G type stars, as well around the K-type and substantially better around M-dwarfs, almost as good as the anoxygenic model.

In Fig. \ref{fig:contours} \textbf{B}, \textbf{C}, and \textbf{D} we show the same information plotted as interpolated \emph{contours} of constant $\nu/\nu_{max}$ as a function of $T_s$ and $a$, for the oxygenic, anoxygenic and hypothetical models respectively. The habitable zone (as defined in \citet{hall2023new}) is shown in green. For oxygenic photosynthesis, the highest photosynthetic rates are confined to G-type stars, with moderate performance across K-types. Both anoxygenic and hypothetical photosynthesis perform consistently across the entire habitable zone.

\subsection{Photosynthetic \textit{difficulty} and optical properties}
\begin{figure*}
    \centering
    {\includegraphics[width=\textwidth]{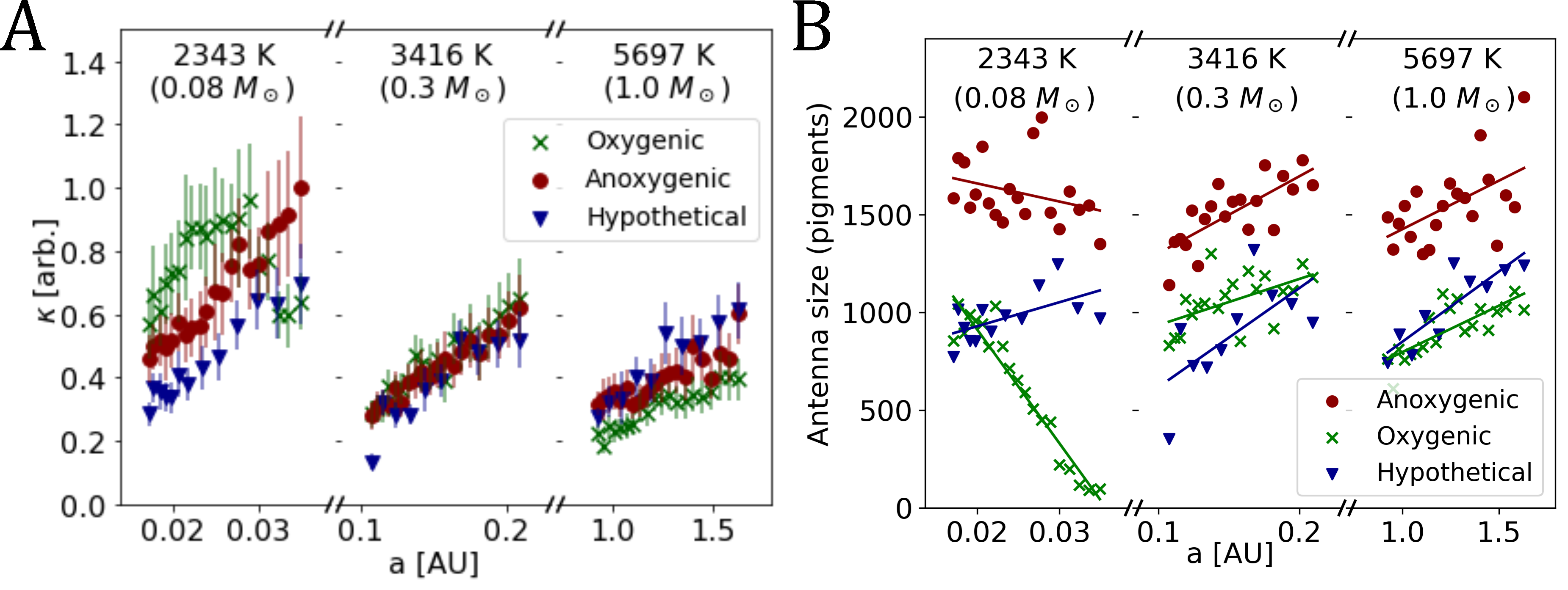}}
     \caption{\textbf{A.} The \emph{relative difficulty} $\kappa$ of different types of photosynthesis around M-, K- and G-stars, with error bars indicating variance across the final population of output by the genetic algorithm. The paradoxical drop in $\kappa$ for the oxygenic mechanism at large $a$ is due to the genetic algorithm failing to find any fit solution. \textbf{B.} The average size of the generated antennae, calculated as $ \left< n_b \sum_{s} N_{s, p} \right> $ across final converged populations for each data point. Note the drop in antenna size for the oxygenic mechanism for the M-type star.}
      \label{fig:costs}
\end{figure*}

Fig. \ref{fig:costs} \textbf{A} plots an estimate of the \emph{difficulty} of performing photosynthesis. The parameter,
\begin{align}
K(a,E_{e,\lambda},\mathcal{M})=\frac{\langle N_p(a,E_{e,\lambda},\mathcal{M})\rangle}{\langle \nu(a,E_{e,\lambda},\mathcal{M})\rangle}
\end{align}
is the average number of antenna pigments needed to produce $\nu=1$ s$^{-1}$. The braces $\langle \ldots\rangle$ indicate an average over the final population produced by the genetic algorithm. We then plot the \emph{relative} difficulty,
\begin{align}
\kappa(a,E_{e,\lambda},\mathcal{M}) = \frac{K(a,E_{e,\lambda},\mathcal{M})}{K_{max}}     
\end{align}
where $K_{max}$ is the highest difficulty across the whole data set. $\kappa$ increases linearly with increasing orbital radius, as the genetic algorithm selects larger antennae to (partially) offset reduced flux. The anoxygenic model experiences a $\sim 20\%$ increase in difficulty around the cool M-dwarf relative to the G-type star. The oxygenic model experiences a $\sim 300\%$ increase, with the paradoxical decrease in $\kappa$ towards the outer edge of the habitable zone due to the evolutionary algorithm failing to find a viable solution at such low flux. Finally, $\kappa$ is largely independent of $T_s$ for the hypothetical model, which experiences half the difficulty of the oxygenic model around the cool M-dwarf.

Fig.~\ref{fig:costs} \textbf{B} shows the average size of the antenna, measured as an average across all converged populations per data point.

\begin{figure}
    {\includegraphics[width=0.48\textwidth]{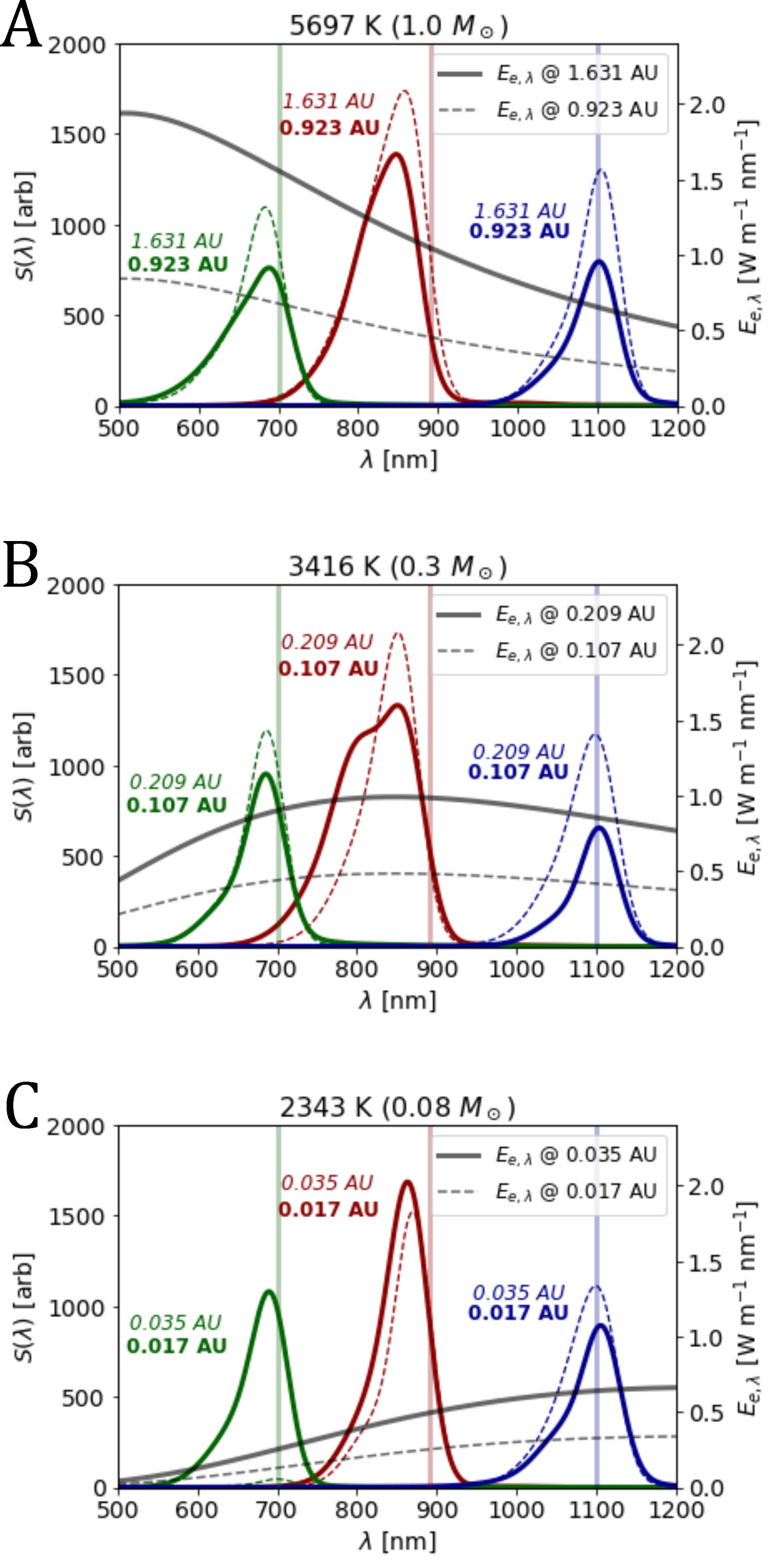}}
     \caption{\textbf{A.} The average absorption spectra, $A_a(\lambda)$, of the evolved photosynthetic models, scaled by antenna size $\langle N_{p}^s \rangle$, at different points across the habitable zone around a G-type star. Green denotes the oxygenic model, red the anoxygenic model, and blue the hypothetical, with thick lines denoting the inner edge of the habitable zone and dashed lines denoting the outer edge. Also shown (in gray) is the incident spectral irradiance, $E_{e,\lambda}(a,T_s)$ at the inner (thick line) and outer (dashed) edges. Finally, the vertical lines indicate the RC absorption wavelengths for each photosynthesis model (700 nm, 890 nm, and 1100 nm for oxygenic, anoxygenic and hypothetical respectively). \textbf{B.} The same as A. but for a K-type star. \textbf{C.} Same as A., but for an M-type star. The genetic algorithm fails to find a workable antenna at the outer edge of the habitable zone, resulting in a tiny antenna and a barely visible $S(\lambda)$.}
      \label{fig:absorption}
\end{figure}

Finally, we can look a the average optical properties of the evolved photosynthetic antennae produced by the genetic algorithm. We define,
\begin{align}
S(\lambda,a,E_{e,\lambda},\mathcal{M}) = \langle N_p^a(a,E_{e,\lambda},\mathcal{M}) \rangle \langle A_a(\lambda,a,E_{e,\lambda},\mathcal{M}) \rangle
\end{align}
as the average \emph{size-weighted} absorption spectrum of the antenna, where $\langle A_a(\lambda,a,E_{e,\lambda},\mathcal{M}) \rangle$ is the average absorption spectrum over the final population. In Fig. \ref{fig:absorption} \textbf{A.} we plot $S(\lambda)$ for a G-type star. Green, red and blue denote the oxygenic, anoxygenic and hypothetical models respectively, and the solid and dashed lines denote $S(\lambda)$ at the inner edge and outer edge of the habitable zone respectively. Also shown is is the flux (light blue) for the same orbital radii, and the vertical lines denote the RC absorption wavelengths, $\lambda_{0,m}$. For each type of photosynthesis, the genetic algorithm yields antennae which absorb in tight bands close to $\lambda_{0,m}$. The absorption profiles are slightly broadened and blue-shifted relative to to the RC absorption $A(\lambda;\lambda_{0,m})$, which reflects the shallow energy funnel structure. Notably, the genetic algorithm does not yield `\emph{black}' antennae that capture all incident flux. As in real photosynthetic systems, this would require a huge antenna with a large number of different types of pigment, which would cost more than any benefit it brought. As $E_{e,\lambda}$ decreases with increasing orbital radius, the genetic algorithm selects for larger antennae.

Fig. \ref{fig:absorption} \textbf{B} shows the same plot but for a K-type star. The broad trends are the same, the only difference being that the evolved antenna for the anoxygenic model has a very broad, two-peak absorption profile. Clearly, the capturing of additional flux is worth the larger antenna, though this advantage disappears at larger orbital radii. 

Finally, Fig. \ref{fig:absorption} \textbf{C} shows the same plot for an M-type star. Both the anoxygenic and hypothetical photosynthesis models benefit from having a larger, more costly antenna. Interestingly, as the orbital radius increases the anoxygenic antenna does not get bigger, simply less blue-shifted. The flux becomes so low that the system does not produce the surplus photosynthetic output needed to build and maintain a larger antenna; the better strategy is simply to focus more on collecting the flux at $~ 900$ nm. The oxygenic model performs adequately at the inner edge of the habitable zone, though it requires a large and costly antenna. At larger orbital distances, the flux becomes so low that building and maintaining any sort of antenna becomes almost impossible. At the outer edge of the habitable zone the antenna absorption profile is barely visible as a dashed curve centred on 700 nm.   
     
\section{Discussion}

Our simulations predict that some form of oxygenic photosynthesis is possible across the entire main sequence habitable zone, though M-dwarfs may select for the hypothetical NIR-driven form rather than the PAR-driven form that evolved on Earth. The Earth-like oxygenic photosynthesis suffers from a sharp decrease in PAR flux at low stellar temperatures/masses, with any antenna structure running into a thermodynamic wall. An antenna large enough to capture the required amount of flux would struggle to transfer that energy efficiently to the reaction centre. 

The hypothetical NIR-driven oxygenic photosynthesis is surprisingly effective, despite the significant kinetic bottleneck of having three photosystems operating in series. This is due to incident flux in the $1000 - 1100$ nm band across the habitable zone being largely independent of stellar temperature/mass. For the coolest M-dwarfs, it out-performs Earth-like oxygenic photosynthesis by at least factor of $2$. Of course, while we have tried to be as agnostic as possible, our models contain some assumed parameters (see Section 1.4 of the Supplementary Material) which could be fine-tuned to make the Earth-like oxygenic mechanism more competitive in M-dwarf light. \citet{battistuzzi2023growth} showed very clearly that cyanobacteria can grow in such light. However, the hypothetical oxygenic mechanism can achieve the same chemical outcome as the Earth-like mechanism while being able to access vastly greater photon flux.  

The simulations also predict that anoxygenic photosynthesis performs well across entire habitable zone, again due to the relatively consistent incident flux in the  $800 - 900$ nm band. It seems to consistently perform better than the two oxygenic mechanisms, delivering a higher rate of $CO_2$ reduction across a wider range of conditions. However, we should not over-interpret this result. There are several external factors that we do not account for, such as the relative abundances of the electron donors or the catalytic complexity of donor oxidation and $CO_2$ reduction. The geochemical reductants ($H_2S$, $H_2$, etc.) that serve as electron donors to Earth's anoxygenic archaea and bacteria are far less abundant globally than $H_2O$. The selection pressure for the evolution of oxygenic photosynthesis on Earth was likely not the abundance of visible light, but the abundance of $H_2O$. A natural extension of this work would be to include the availabilities of these required reductants in the model, and investigate the effect of changes in relative availability on the preferred mode of photosynthesis.

Lastly, we can propose some constraints on potential reflectance biosignatures from different photosynthetic mechanisms. While our simulations don't predict the evolutionary origin of photosynthesis, or the timescales involved, we can predict a sequence of events. 

Firstly, a single functional photosystem has to evolve before multiple photosystems can be fused into complex photo-redox chains. This means that the first photosynthetic mechanism is unlikely to be oxygenic. The free energy change associated with the one-electron pseudo-reaction,
\begin{align}
    \frac{1}{2}H_2O + \frac{1}{4}CO_2 \rightarrow \frac{1}{4}O_2 + \frac{1}{4}CH_2O + \frac{1}{4}H_2O
\end{align}
is only
\begin{align}
 \Delta G_{red}(CH_2O) - \Delta G_{ox}(H_2O) \sim 1.25 \text{ eV} \sim 990 \text{ nm}  
\end{align}
However, as we have argued, an excess of energy is needed to drive charge separation, donor oxidation and electron transfer. If we assume similar energy gaps ($\sim 15$ $k_BT$) to those in Fig. \ref{fig:levels}, then this would need a photon of $\lambda \sim 510$~nm. Since an antenna system can only ever harvest light \emph{bluer} than the RC absorption, a hypothetical one-photon oxygenic mechanism would be extremely limited in the light it could utilize. Moreover, photosynthesis has to evolve as an incremental adaptation of some already established chemosynthetic process, which is very unlikely to involve the extremely endergonic processes of water oxidation. Reflectance edges at $800 - 900$ nm may therefore be a universal marker of life in the universe \citep[as argued by][]{10.1093/mnras/stae601}. 

Once established, the first anoxygenic photosynthetic process could undergo divergent evolution, with variations in the ionization potential of the RC ($I_m = G(P_m^+)-G(P_m)$) allowing for variation in the electron donor. Oxygenic photosynthesis could then evolve as a fusion of two (or three) of these slightly different anoxygenic systems \citep[see][]{schmidt2021}. This is almost certainly how oxygenic photosynthesis evolved on Earth, with Photosystem II and Photosystem I evolving from the Type II and Type I anoxygenic reaction centres. 

Our simulations suggest that, for K and G-type stars, an Earth-like oxygenic mechanism provides a clear evolutionary advantage. Therefore, the canonical $700 - 800$ nm Vegetation Red Edge (VRE) could be expected for more established biospheres on older planets. Around M-dwarfs, we could still expect the evolution of oxygenic photosynthesis though it may be the NIR-driven variant. The additional redox complexity of three rather than two photosystems is a minor difference compared to the complexity of evolving a functional photosystem in the first place. For exoplanets orbiting M-dwarfs we should therefore also look for reflectance biosignatures in the $1000 - 1100$ nm band \citep[see][]{lehmer2021peak}.

We also note that we have neglected the effect of atmospheric attenuation in the current work, since adding a general consideration of possible atmospheric compositions increases the parameter space and hence the computational complexity significantly. While this is of course an approximation, previous work~\citep{chitnavis2024optimizing} has suggested that fine spectral features in the incident flux should not have qualitative effects on antenna composition, due to the wide absorption profile of biological chromophores. However, further work which combines the current antenna model with a detailed description of atmospheric composition will be necessary and we welcome collaborations with groups who wish to investigate evolution on exoplanets under specific atmospheric conditions.

\section*{Data availability statement}

The data used to create the plots in this article are listed in tables in the Supplementary Material. The code used to generate the data in this work is available on github at \url{https://github.com/QMUL-DuffyLab/gala} and all data generated will be available on Zenodo at the time of publication.

\section*{Author contributions}
CG, CH, AR, TH, EG and CDPD devised the project. CDPD, SS and KSR developed the redox models of the oxygenic, anoxygenic and hypothetical photosystems, with SS providing the justifications for timescales. CDPD and CG developed the equations of motion in discussion with SS. CH provided the calculations of spectral irradiances and the definitions of the habitable zone. The genetic algorithm was developed by CG, who carried out the bulk of the data generation and analysis. CG also administers the project github and Zenodo repositories. CH, AR, TH and EG provided the astronomical context for the results, including developing the contour plots. CDPD, CG, and CH wrote the manuscript, with editing by SS, KSR, AR, TH and EG.   

\section*{Acknowledgements}
CG, CDPD, EG and TJH would like to acknowledge the support of the Leverhulme Trust Project Grant (RPG-2023-096). 

\bibliographystyle{mnras}
\bibliography{phz}



\begin{appendix}
\section{Photosynthetic Kinetic Equations}

As in \citet{chitnavis2024optimizing} and \citet{gray2025predicting}, we adopt a stochastic model based on the time-evolution of joint probabilities associated with different configuration of the whole supersystem.

\subsection{Joint Probabilities for the Instantaneous Configuration Photosynthetic Supersystem}

The photosynthetic supersystem is a linear chain of photosystems, starting with the donor-oxidizing photosytems $PS_{ox}$ and ending in the $CO_2$-reducing photosystem, $PS_r$ 
\begin{align}
    PS_{ox} \rightarrow \ldots \rightarrow PS_m \rightarrow \ldots PS_r
\end{align}
Each $PS_m$ contains the low energy pigment, $P_m$, and the redox cofactor, $T_m$, 
\begin{align}
    PS_m = P_m + T_m
\end{align}
and so each photosystem can be in one of five electronic/redox states: 
\begin{enumerate}
    \item{Ground state: $\text{P}_m+T_m$}
    \item{Excited: $[\text{P}_m]^*+T_m$}
    \item{Charge Separated: $[\text{P}_m]^+ +[T_m]^-$}
    \item{Reduced}: $\text{P}_m +[T_m]^-$
     \item{Oxidized: $[\text{P}_m]^+ +T_m$} 
\end{enumerate}
We therefore define the occupation numbers,
\begin{align}
    n_x^m&=0,1\\
    n_{cs}^m&=0,1\\
    n_-^m&=0,1\\
    n_+^m&=0,1
\end{align}
where $n_x^m$, $n_{cs}^m$, $n_-^m$, and $n_+^m$ denote the occupancies of the excited, charge separated, reduced and oxidized states respectively. We can then define the vector, 
\begin{align}
    \mathbf{n}^m=\left(n_x^m,n_{cs}^m,n_-^m,n_+^m\right)
\end{align}
We note that the five states are mutually exclusive, meaning we can define label the specific vectors,
\begin{align}
\mathbf{gs}^m &= (0,0,0,0) = P_m+T_m\\
 \mathbf{ex}^m &= (1,0,0,0)= P_m^*+T_m\\
 \mathbf{cs}^m &= (0,1,0,0) = P_m^++T_m^-\\
 \mathbf{r}^m &= (0,0,1,0)=P_m+T_m^-\\
\mathbf{ox}^m &= (0,0,0,1) = P_m^++T_m
\end{align}

The photosystems are coupled to a shared antenna systems which is an assembly of $N^a_s$ subunits, each of which contains $N^a_{s,p}$ pigments. We can then define the exciton occupancies,
\begin{align}
    n^a_{s} = 0,1, \ldots, N^a_{s,p}
\end{align}
where $n^a_s$ denotes the number of excitons in antenna subunit $s$. We can therefore define the vector,
\begin{align}
 \mathbf{n}^a = (n^a_1, n^a_2,\ldots,n^a_s,\ldots,n^a_{N_{s}})   
\end{align}
As in \citet{chitnavis2024optimizing} and \citet{gray2025predicting} we invoke the \emph{single excitation approximation}, which is the assumption that there is never more than one exciton in the antenna at any time,
\begin{align}
\sum_s n^a_s = 0 \text{ or } 1    
\end{align}
This is fully justified by the fact that photon absorption, even in bright light, is generally much slower than de-excitation processes. We therefore define the vectors,
\begin{align}
\mathbf{0}^a &= (0,0,\ldots,0,\ldots,0)\\
\mathbf{1}_1^a &= (1,0,\ldots,0,\ldots,0)\\
\mathbf{1}_2^a &= (0,1,\ldots,0,\ldots,0)\\
\vdots\notag\\
\mathbf{1}_s^a &= (0,0,\ldots,1,\ldots,0)\\
\vdots\notag\\
\mathbf{1}_{N_s^a}^a &= (0,0,\ldots,0,\ldots,1)
\end{align}

The full set of occupancies defines the instantaneous configuration of our photosynthetic supersystem. For each configuration we can assign a time-dependent probability,
\begin{align}
P(t)=P_{\mathbf{n}^a,\mathbf{n}^{ox},\ldots,\mathbf{n}^m,\ldots,\mathbf{n}^r}(t)
\end{align}
normalized so that,
\begin{align}
\sum_{\mathbf{n}^a}\sum_{\mathbf{n}^{ox}}\ldots \sum_{\mathbf{n}^m}\ldots\sum_{\mathbf{n}^r}
P(t)=1
\end{align}

\subsection{The Equations of Motion}

The supersystem transitions between different configurations via to various photochemical processes,
\begin{align}\label{eq:dPdtapp}
\begin{split}    
\frac{d}{dt}P(t)&=\left[\frac{d}{dt}P(t)\right]_{\gamma}+\left[\frac{d}{dt}P(t)\right]_{ET}+\left[\frac{d}{dt}P(t)\right]_{D}\\
&+\left[\frac{d}{dt}P(t)\right]_{CS}+\left[\frac{d}{dt}P(t)\right]_{e^-}+\left[\frac{d}{dt}P(t)\right]_{cyc}\\
&+\left[\frac{d}{dt}P(t)\right]_{HA}+\left[\frac{d}{dt}P(t)\right]_{CO_2}
\end{split}
\end{align}
We shall list the equations for each of these processes below. To reduce notational complexity we will only display \emph{relevant indices}. 

\subsubsection{The photoexcitation kinetics}

The first term in Eqn. (\ref{eq:dPdtapp}) describes photon absorption. The two paired equations,  
\begin{align}
\left[\frac{d}{dt}P_{\mathbf{0}^a}(t)\right]_\gamma &= -\sum_{s} \gamma_{s}  P_{\mathbf{0}^a}(t)\\
\left[\frac{d}{dt}P_{\mathbf{1}^a_{s'}}(t)\right]_\gamma &=  \gamma_{s'}  P_{\mathbf{0}^a}(t)
\end{align}
describe the creation of an exciton in a subunit of the antenna,
\begin{align}
\left(\mathbf{0}^a,\mathbf{n}^{ox}\ldots,\mathbf{n}^m,\ldots,\mathbf{n}^r\right)\xrightarrow{\gamma}\left(\mathbf{1}^a_{s'},\mathbf{n}^{ox}\ldots,\mathbf{n}^m,\ldots,\mathbf{n}^r\right)  
\end{align}
where $\gamma_s$ is the excitation rate for subunit $s$. As in the main article, the excitation rate is given by,
\begin{align}
\gamma_s = \sigma_{s,p} N_{s,p}^a \int_0^\infty d\lambda \frac{\lambda}{hc} E_{e,\lambda}A_s(\lambda,\lambda_{0,s})
\end{align}
where $\sigma_{s,p}$ is the integrated absorption cross-section of a single pigment in subunit $s$, $N_{s,p}^a$ is the number of pigments in subunit $s$, $E_{e,\lambda}$ is the spectral irradiance, and $A_s(\lambda,\lambda_{0,s})$ is the normalized absorption spectrum of subunit $s$. 

The pigment within the reaction centres, $Pm$, can also be excited via  photon absorption,
\begin{align}
\left[\frac{d}{dt}P_{\mathbf{0}^a,\mathbf{gs}^m}(t)\right]_\gamma &= - \gamma_{m}  P_{\mathbf{0}^a,\mathbf{gs}^m}(t)\\
\left[\frac{d}{dt}P_{\mathbf{0}^a,\mathbf{ex}^m}(t)\right]_\gamma &=  \gamma_{m}  P_{\mathbf{0}^a,\mathbf{gs}^m}(t)
\end{align}
which describe the flow of probability,
\begin{align}
\left(\mathbf{0}^a,\mathbf{n}^{ox}\ldots,\mathbf{gs}^m,\ldots,\mathbf{n}^r\right)\xrightarrow{\gamma}\left(\mathbf{0}^a,\mathbf{n}^{ox}\ldots,\mathbf{ex}^m,\ldots,\mathbf{n}^r\right)     
\end{align}
The photoexcitation rate $\gamma_m$ is described in an identical manner to that of the antenna,
\begin{align}
\gamma_m = \sigma_{m} \int_0^\infty d\lambda \frac{\lambda}{hc} E_{e,\lambda}A_m(\lambda,\lambda_{0,m})
\end{align}
but for the fact there is only a single pigment within $PS_m$. 

\subsubsection{Exciton transfer kinetics}

The transfer of an exciton between sub-units $s$ and $s'$,
\begin{align}
\left(\mathbf{1}_s^a,\mathbf{n}^{ox}\ldots,\mathbf{n}^m,\ldots,\mathbf{n}^r\right)\xrightleftharpoons[k_{s\rightarrow s'}]{k_{s'\rightarrow s}} \left(\mathbf{1}_{s'}^a,\mathbf{n}^{ox}\ldots,\mathbf{n}^m,\ldots,\mathbf{n}^r\right)
\end{align}
is described by,
\begin{align}
\left[\frac{d}{dt}P_{\mathbf{1}_s^a}(t)\right]_{ET}&=-k_{s\rightarrow s'}P_{\mathbf{1}_s^a}(t) + k_{s'\rightarrow s}P_{\mathbf{1}_{s'}^a}(t)\\
\left[\frac{d}{dt}P_{\mathbf{1}_{s'}^a}(t)\right]_{ET}&=k_{s\rightarrow s'}P_{\mathbf{1}_s^a}(t) - k_{s'\rightarrow s}P_{\mathbf{1}_{s'}^a}(t)    
\end{align}
where $k_{s\rightarrow s'}$ and $k_{s'\rightarrow s}$ are the forward and backward exciton transfer rate constants. The rate constants are given by,
\begin{align}\label{eq:k_hop}
k_{s\rightarrow s'}= a_{s,s'} \frac{1}{\tau_{hop}} D(\lambda_{0,s},\lambda_{0,s'}) F\left(\Delta G_{s\rightarrow s'}\right)    
\end{align}
where $\tau_{hop}$ is the characteristic time constant for the process, which is then subject to several modifiers. The first is $a_{s,s'}$, which is an element of the \emph{adjacency matrix},
\begin{align}
    a_{s,s'}\begin{cases}
        1 &\text{for } s \text{ and } s' \text{ connected}\\
        0 &\text{otherwise}
    \end{cases}
\end{align}
which determines the overall topology of the antenna. Essentially, energy transfer is only possible between connected/neighbouring subunits. 

The next is the spectral overlap integral,
\begin{align}
D_{s,s'}(\lambda_{0,s},\lambda_{0,s'})=\int_0^\infty d\lambda F_{s}(\lambda;\lambda_{0,s})A_{s'}(\lambda;\lambda_{0,s'})
\end{align}
where $F_s(\lambda)$ and $A_{s'}(\lambda)$ are the fluorescence spectrum of the donor, $s$, and the absorption spectrum of the acceptor, $s'$, respectively. Essentially, energy transfer between subunits with different excitation energies is only possible because vibrational fluctuations occasionally bring them into energetic \emph{resonance}. The degree to which which this happens is measured by the spectral overlap. 

The final modifier enforces detailed balance,
\begin{align}
F\left(\Delta G_{s\rightarrow s'}\right)&=\begin{cases}
1 & \text{if } \Delta G_{s\rightarrow s'} \leq 0\\
\exp \left(-\frac{\Delta G_{s\rightarrow s'}}{k_BT_p}\right) & \text{if } \Delta G_{s\rightarrow s'} > 0
\end{cases}    
\end{align}
where $\Delta G_{s\rightarrow s'}$ is the free energy change associated with the transfer of an exciton from $s$ to $s'$, $k_B$ is Boltzmann's constant, and $T_p$ is the average surface temperature on the planet. This ensures that processes that increase free energy are subject to a thermodynamic penalty. 

By the same logic, energy transfer between the antenna and one of the reaction centres,
\begin{align}
\left(\mathbf{1}^a_{s'},\mathbf{n}^{ox},\ldots,\mathbf{gs}^m,\ldots,\mathbf{n}^{r}\right) \xrightleftharpoons[k_{s'\rightarrow m}]{k_{m\rightarrow s'}} \left(\mathbf{0}^a,\mathbf{n}^{ox},\ldots,\mathbf{ex}^m,\ldots,\mathbf{n}^{r}\right)    
\end{align}
is given by,
\begin{align}
\begin{split}    
\left[\frac{d}{dt}P_{\mathbf{1}_{s'}^a,\mathbf{gs}^m}(t)\right]_{ET}&=-k_{s'\rightarrow m} P_{\mathbf{1}_{s'}^a,\mathbf{gs}^m}(t)\\ 
&+ k_{m\rightarrow s'} P_{\mathbf{0}^a,\mathbf{ex}^m}(t)    
\end{split}\\
\begin{split}    
\left[\frac{d}{dt}P_{\mathbf{0}^a}\mathbf{ex}^m(t)\right]_{ET}&=\sum_{s}k_{s\rightarrow m} P_{\mathbf{1}_{s}^a,\mathbf{gs}^m}(t)\\ 
&- \sum_{s}k_{m\rightarrow s} P_{\mathbf{0}^a,\mathbf{ex}^m}(t)         
\end{split}
\end{align}
where the rate constants $k_{s'\rightarrow m}$ and $k_{m \rightarrow s'}$ are defined in exactly the same way as $k_{s\rightarrow s'}$. 

\subsubsection{Excitation dissipation}

The thrid term in Eqn. (\ref{eq:dPdtapp}) characterizes the dissipation of excitons via decay processes such as fluorescence, internal conversion non-radiative decay), etc. Since we are not interested in these processes individually, we collect them all into a single decay term,
\begin{align}
\left[\frac{d}{dt}P_{\mathbf{1}_{s'}^a}(t)\right]_{D} = - k_{diss} P_{\mathbf{1}_{s'}^a}(t)\\    
\left[\frac{d}{dt}P_{\mathbf{0}^a}(t)\right]_{D} =  k_{diss} \sum_s P_{\mathbf{1}_{s}^a}(t)
\end{align}
where $k_{diss}$ is the rate constant for dissipation, which is assumed to be identical for all subunits. This equation characterize dissipation in the antenna,
\begin{align}
\left(\mathbf{1}_{s'}^a,\mathbf{n}^{ox},\ldots,\mathbf{n}^m,\ldots,\mathbf{n}^r\right)\xrightarrow{k_{diss}} \left(\mathbf{0}^a,\mathbf{n}^{ox},\ldots,\mathbf{n}^m,\ldots,\mathbf{n}^r\right)    
\end{align}
The reaction centre pigments can also decay in this manner,
\begin{align}
\left(\mathbf{0}^a,\mathbf{n}^{ox},\ldots,\mathbf{ex}^m,\ldots,\mathbf{n}^r\right)\xrightarrow{k_{diss}} \left(\mathbf{0}^a,\mathbf{n}^{ox},\ldots,\mathbf{gs}^m,\ldots,\mathbf{n}^r\right)    
\end{align}
which is characterized by,
\begin{align}
\left[\frac{d}{dt}P_{\mathbf{ex}^m}(t)\right]_{D} = - k_{diss} P_{\mathbf{ex}^m}(t)\\    
\left[\frac{d}{dt}P_{\mathbf{gs}^m}(t)\right]_{D} =  k_{diss}  P_{\mathbf{ex}^m}(t)    
\end{align}
where we assume the same dissipation rate, $k_{diss}$.

\subsubsection{Charge separation kinetics}

Charge separation in the reaction centres,
\begin{align}
\left(\mathbf{n}^a,\mathbf{n}^{ox},\ldots,\mathbf{ex}^m,\ldots,\mathbf{n}^r\right)\xrightarrow{k_{cs}}\left(\mathbf{n}^a,\mathbf{n}^{ox},\ldots,\mathbf{cs}^m,\ldots,\mathbf{n}^r\right)    
\end{align}
is characterized by,
\begin{align}
\left[\frac{d}{dt}P_{\mathbf{ex}^m}(t)\right]_CS &= -k_{cs} P_{\mathbf{ex}^m}(t)\\     
\left[\frac{d}{dt}P_{\mathbf{cs}^m}(t)\right]_CS &= k_{cs} P_{\mathbf{ex}^m}(t)
\end{align}
where $k_{cs}$ is the charge separation rate. Note, the reverse process, in which the exciton reforms from the charge separated state, negligible (see the next section). 

\subsubsection{Linear electron transfer between photosystems}

Linear electron transfer involves the transfer of an electron from one photosystem, $PS_m$, to the next one along the chain, $PS_{m+1}$. So we don't enforce a particular time-ordering on the various redox processes, we have to consider several possibilities. The first involves both $PS_m$ and $PS_{m+1}$ starting in the charge separated state and ending with $PS_m$ oxidized and $PS_{m+1}$ reduced, 
\begin{align}
\left[P_m^+ + T_m^-\right] + \left[P_{m+1}^+ + T_{m+1}^-\right] &\rightarrow  \left[P_m^+ + T_m\right] + \left[P_{m+1} + T_{m+1}^-\right]   
\end{align}
In terms of indices this is,
\begin{align}
&\left(\mathbf{n}^x,\mathbf{n}^{ox},\ldots,\mathbf{cs}^m,\mathbf{cs}^{m+1},\ldots,\mathbf{n}^r\right)\\
\xrightarrow{k_{e^-}} &\left(\mathbf{n}^x,\mathbf{n}^{ox},\ldots,\mathbf{ox}^m,\mathbf{r}^{m+1},\ldots,\mathbf{n}^r\right) \nonumber
\end{align}
with the kinetics described by,
\begin{align}
\left[\frac{d}{dt}P_{\mathbf{cs}^m,\mathbf{cs}^{m+1}}(t)\right]_{e^-} &= -k_{e^-} P_{\mathbf{cs}^m,\mathbf{cs}^{m+1}}(t)\\   
\left[\frac{d}{dt}P_{\mathbf{ox}^m,\mathbf{r}^{m+1}}(t)\right]_{e^-} &= k_{e^-} P_{\mathbf{cs}^m,\mathbf{cs}^{m+1}}(t)   
\end{align}

It is also possible that the process starts with $PS_m$ in the charge separated state and $PS_{m+1}$ in the oxidized state,
\begin{align}
&\left[P_m^+ + T_m^-\right] + \left[P_{m+1}^+ + T_{m+1}\right]\\
\rightarrow &\left[P_m^+ + T_m\right] + \left[P_{m+1} + T_{m+1}\right] \nonumber 
\end{align}
or,
\begin{align}
&\left(\mathbf{n}^x,\mathbf{n}^{ox},\ldots,\mathbf{cs}^m,\mathbf{ox}^{m+1},\ldots,\mathbf{n}^r\right)\\
\xrightarrow{k_{e^-}}&\left(\mathbf{n}^x,\mathbf{n}^{ox},\ldots,\mathbf{ox}^m,\mathbf{gs}^{m+1},\ldots,\mathbf{n}^r\right) \nonumber
\end{align}
with kinetics,
\begin{align}
\left[\frac{d}{dt}P_{\mathbf{cs}^m,\mathbf{ox}^{m+1}}(t)\right]_{e^-} &= -k_{e^-} P_{\mathbf{cs}^m,\mathbf{ox}^{m+1}}(t)\\   
\left[\frac{d}{dt}P_{\mathbf{ox}^m,\mathbf{gs}^{m+1}}(t)\right]_{e^-} &= k_{e^-} P_{\mathbf{cs}^m,\mathbf{ox}^{m+1}}(t)   
\end{align}

We could also start with $PS_m$ in the reduced state and $PS_{m+1}$ in the charge separated state,
\begin{align}
\left[P_m + T_m^-\right] + \left[P_{m+1}^+ + T_{m+1}^-\right] &\rightarrow  \left[P_m + T_m\right] + \left[P_{m+1} + T_{m+1}^-\right]       
\end{align}
or,
\begin{align}
&\left(\mathbf{n}^x,\mathbf{n}^{ox},\ldots,\mathbf{r}^m,\mathbf{cs}^{m+1},\ldots,\mathbf{n}^r\right)\\
\xrightarrow{k_{e^-}} &\left(\mathbf{n}^x,\mathbf{n}^{ox},\ldots,\mathbf{gs}^m,\mathbf{r}^{m+1},\ldots,\mathbf{n}^r\right) \nonumber
\end{align}
with kinetics,
\begin{align}
\left[\frac{d}{dt}P_{\mathbf{r}^m,\mathbf{cs}^{m+1}}(t)\right]_{e^-} &= -k_{e^-} P_{\mathbf{r}^m,\mathbf{cs}^{m+1}}(t)\\   
\left[\frac{d}{dt}P_{\mathbf{gs}^m,\mathbf{r}^{m+1}}(t)\right]_{e^-} &= k_{e^-} P_{\mathbf{r}^m,\mathbf{cs}^{m+1}}(t)   
\end{align}

Finally, we could start with $PS_m$ in the reduced state and $PS_{m+1}$ in the oxidized states,
\begin{align}
&\left[P_m + T_m^-\right] + \left[P_{m+1}^+ + T_{m+1}\right]\\
\rightarrow &\left[P_m + T_m\right] + \left[P_{m+1} + T_{m+1}\right]       
\end{align}
or,
\begin{align}
&\left(\mathbf{n}^x,\mathbf{n}^{ox},\ldots,\mathbf{r}^m,\mathbf{ox}^{m+1},\ldots,\mathbf{n}^r\right)\\
\xrightarrow{k_{e^-}} &\left(\mathbf{n}^x,\mathbf{n}^{ox},\ldots,\mathbf{gs}^m,\mathbf{gs}^{m+1},\ldots,\mathbf{n}^r\right) \nonumber
\end{align}
with kinetics,
\begin{align}
\left[\frac{d}{dt}P_{\mathbf{r}^m,\mathbf{ox}^{m+1}}(t)\right]_{e^-} &= -k_{e^-} P_{\mathbf{r}^m,\mathbf{ox}^{m+1}}(t)\\   
\left[\frac{d}{dt}P_{\mathbf{gs}^m,\mathbf{gs}^{m+1}}(t)\right]_{e^-} &= k_{e^-} P_{\mathbf{r}^m,\mathbf{ox}^{m+1}}(t)   
\end{align}

We note that for all of these processes, we assume a common rate constant $k_{lin}$. In reality, the rate constant will almost certainly depend, to some extent, on the redox state of the donor and acceptor, though this introduces a large number of free parameters and is difficult to generalize. 

\subsubsection{Cyclic electron flow}

In our anoxygenic model we assume that chemical energy is generated by cyclic electron flow about the single photosystem, $PS_{an}$,
\begin{align}
 P_{an}^++T_{an}^- \xrightarrow{k_{cyc}} P_{an} + T_{an} + \text{Energy}   
\end{align}
In terms of indices this is,
\begin{align}
\left(\mathbf{n}^a,\mathbf{cs}^{an}\right)\xrightarrow{k_{cyc}}\left(\mathbf{n}^a,\mathbf{gs^{an}}\right)    
\end{align}
whith the kinetics given by,
\begin{align}
\left[\frac{d}{dt}P_{\mathbf{cs}^{an}}(t)\right]_{cyc}&=-k_{cyc}P_{\mathbf{cs}^{an}}(t)\\    
\left[\frac{d}{dt}P_{\mathbf{gs}^{an}}(t)\right]_{cyc}&=k_{cyc}P_{\mathbf{cs}^{an}}(t)\\    
\end{align}
where $k_{cyc}$ is the associated rate constant. We note that these equation are almost identical to those that would describe the processes of non-radiative charge recombination, though this is due to the course grained nature of the model. Real cyclic electron flow would likely involve intermediate electron carriers, while recombination is a direct processes. The important distinction is that recombination does not produce usable chemical energy and is extremely slow due to a strong evolutionary pressure to avoid waste processes. 

Another point to note is that, in Earth organisms, cyclic electron flow (about PSI) also occurs in oxygenic photosynthesis \citep{nawrocki2019mechanism}. Since the amount of cyclic depends on environmental conditions and varies from species to species, we neglect it in our model. 

\subsubsection{Donor oxidation kinetics}

The mechanics of donor oxidation are treated in a very course-grained way, assuming that we have an infinite supply of $HA$. As with linear electron transfer, there are more than one pathways for this process,
\begin{align}
P_{ox}^++T_{ox}^- &\xrightarrow{k_{ox}} P_{ox} + T_{ox}^-\\ 
P_{ox}^++T_{ox} &\xrightarrow{k_{ox}} P_{ox} + T_{ox}
\end{align}
In terms of indices this is,
\begin{align}
\left(\mathbf{n}^a,\mathbf{cs}^{ox},\ldots,\mathbf{n}^m,\ldots,\mathbf{n}^r\right)&\xrightarrow{k_{ox}}\left(\mathbf{n}^a,\mathbf{r}^{ox},\ldots,\mathbf{n}^m,\ldots,\mathbf{n}^r\right)\\    
\left(\mathbf{n}^a,\mathbf{ox}^{ox},\ldots,\mathbf{n}^m,\ldots,\mathbf{n}^r\right)&\xrightarrow{k_{ox}}\left(\mathbf{n}^a,\mathbf{gs}^{ox},\ldots,\mathbf{n}^m,\ldots,\mathbf{n}^r\right)
\end{align}
respectively, with the kinetics determined by,
\begin{align}
\left[\frac{d}{dt}P_{\mathbf{cs}^{ox}}(t)\right]_{HA}&=-k_{ox} P_{\mathbf{cs}^{ox}}(t)\\    
\left[\frac{d}{dt}P_{\mathbf{r}^{ox}}(t)\right]_{HA}&=k_{ox} P_{\mathbf{cs}^{ox}}(t)
\end{align}
and,
\begin{align}
\left[\frac{d}{dt}P_{\mathbf{ox}^{ox}}(t)\right]_{HA}&=-k_{ox} P_{\mathbf{ox}^{ox}}(t)\\    
\left[\frac{d}{dt}P_{\mathbf{gs}^{ox}}(t)\right]_{HA}&=k_{ox} P_{\mathbf{ox}^{ox}}(t)
\end{align}
respectively. Again, for model simplicity we assume a single rate constant, $k_{ox}$, for both of these pathways.

\subsubsection{$CO_2$ reduction kinetics}

Finally, we adopt a similar course-grained representation of $CO_2$ reduction, which can occur through via two pathways,
\begin{align}
 P_r^+ + T_r^- &\xrightarrow{k_{red}} P_{r}^+ + T_r\\
 P_r + T_r^- &\xrightarrow{k_{red}} P_{r} + T_r
\end{align}
In terms of indices these are,
\begin{align}
\left(\mathbf{n}^a,\mathbf{n}^{ox},\ldots,\mathbf{n}^{m},\ldots,\mathbf{cs}^r\right) &\xrightarrow{k_{red}} \left(\mathbf{n}^a,\mathbf{n}^{ox},\ldots,\mathbf{n}^{m},\ldots,\mathbf{ox}^r\right)\\   
\left(\mathbf{n}^a,\mathbf{n}^{ox},\ldots,\mathbf{n}^{m},\ldots,\mathbf{r}^r\right) &\xrightarrow{k_{red}} \left(\mathbf{n}^a,\mathbf{n}^{ox},\ldots,\mathbf{n}^{m},\ldots,\mathbf{gs}^r\right)
\end{align}
with the kinetics given by,
\begin{align}
\left[\frac{d}{dt}P_{\mathbf{cs}^r}(t)\right]_{CO_2}&=-k_{red} P_{\mathbf{cs}^r}(t)\\    
\left[\frac{d}{dt}P_{\mathbf{ox}^r}(t)\right]_{CO_2}&=k_{red} P_{\mathbf{cs}^r}(t)    
\end{align}
and,
\begin{align}
\left[\frac{d}{dt}P_{\mathbf{r}^r}(t)\right]_{CO_2}&=-k_{red} P_{\mathbf{r}^r}(t)\\    
\left[\frac{d}{dt}P_{\mathbf{gs}^r}(t)\right]_{CO_2}&=k_{red} P_{\mathbf{r}^r}(t)    
\end{align}

\subsection{Solving for the $CO_2$ reduction rate}

The full set of differential equations derived about can be represented in vector form as,
\begin{align}
\frac{d}{dt}\mathbf{P}(t) = \mathbf{K} \mathbf{P}(t)    
\end{align}
where,
\begin{align}
\mathbf{P}(t) = \begin{pmatrix}
& P_{\mathbf{0}^a,\mathbf{gs}^{ox},\ldots,\mathbf{gs}^m,\ldots,\mathbf{gs}^r}(t)\\
& P_{\mathbf{1}_{1}^a,\mathbf{gs}^{ox},\ldots,\mathbf{gs}^m,\ldots,\mathbf{gs}^r}(t)\\
&\vdots
\end{pmatrix}    
\end{align}
is the vector of probabilities for all possible configurations, and $\mathbf{K}$ is the \emph{transfer matrix} that encodes the kinetics. We then then use a non-negative least squares algorithm to find the equilibrium probabilities,
\begin{align}
\mathbf{K} \mathbf{P}^{eq}=0    
\end{align}
subject to the constraint that,
\begin{align}
\sum_i P^{eq}_i = 1    
\end{align}
where $P_i^{eq}$ is the $i^\text{th}$ element of $\mathbf{P}^{eq}$. From the equilibrium populations we can then calculate observable quantities such as the $CO_2$ reduction rate,
\begin{align}
 \nu = k_{red} \sum_{\mathbf{n}^a}\sum_{\mathbf{n}^{ox}}\ldots\sum_{\mathbf{n}^m}\ldots \Big(P^{eq}_{\mathbf{n}^a,\mathbf{n}^{ox},\ldots,\mathbf{n}^{m},\ldots,\mathbf{cs}^r}+P^{eq}_{\mathbf{n}^a,\mathbf{n}^{ox},\ldots,\mathbf{n}^{m},\ldots,\mathbf{r}^r}\Big)   
\end{align}
For the three photosynthesis models considered in the main articles, anoxygenic, oxygenic and hypothetical, the $CO_2$ rates are given by,
\begin{align}
 \nu(\text{anoxygenic}) &= k_{red} \sum_{\mathbf{n}^a} \Big(P^{eq}_{\mathbf{n}^a,\mathbf{cs}^an}+P^{eq}_{\mathbf{n}^a,\mathbf{r}^an}\Big)\\
\nu(\text{oxygenic}) &= k_{red} \sum_{\mathbf{n}^a}\sum_{\mathbf{n}^{ox}}\Big(P^{eq}_{\mathbf{n}^a,\mathbf{n}^{ox}\mathbf{cs}^r}+P^{eq}_{\mathbf{n}^a,\mathbf{r}^r}\Big)\\
\nu(\text{hypothetical}) &= k_{red} \sum_{\mathbf{n}^a}\sum_{\mathbf{n}^{ox}}\sum_{\mathbf{n}^i} \Big(P^{eq}_{\mathbf{n}^a,\mathbf{n}^{ox},\mathbf{n}^{i},\mathbf{cs}^r}+P^{eq}_{\mathbf{n}^a,\mathbf{n}^{ox},\mathbf{n}^{i},\mathbf{r}^r}\Big)   
\end{align}
respectively. 

\subsection{Estimation of rate constants}

The goal of this work is not to precisely reproduce the dynamics of the photosynthetic light reactions in any particular organism, but to derive a robust, general model based on basic (and therefore \emph{universal}) physical/chemical arguments. We will therefore define the various rate constants up to their order of magnitude based on the following assumptions.

\subsubsection{The 'pigments' are most likely $\pi$-conjugated organic molecules}

The photon absorption rates, $\gamma_s$ and $\gamma_m$ depend on the integrated absorption cross-sections $\sigma_s$ and $\sigma_E^m$. We assume these are of the order of
\begin{align}
    \sigma_s=\sigma_E^m\sim 10^{-20} \text{ m}^2
\end{align}, 
which is typical of organic such molecules \citep{Noy_Chla_sigma}. 

We also assume that for any photosynthetic organism, there would be a strong selection pressure to minimize exciton dissipation. Therefore, we assume,  
\begin{align}
\frac{1}{k_\text{diss}}\sim 1 \text{ ns}
\end{align}
which is the typical excitation lifetime of fluorescent molecules \citep{lakowicz2006principles}. 

\subsubsection{Exciton transfer in the antenna is fast}

There will be a strong selection pressure for light-harvesting to efficient, meaning a high pigment density and fast energy transfer. We assume that \emph{intra}-subunit exciton transfer is extremely fast, with exciton equilibration occurring effectively instantly. \emph{Inter}-subunit transfer will be slower, since the distances are larger, though still much faster than exciton dissipation. The exciton transfer rates as defined in Eqn. (\ref{eq:k_hop}) depend on a characteristic timescale $\tau_{hop}$, whihc we assume is o fthe order,
\begin{align}
\tau_{hop} \sim 10 \text{ ps}    
\end{align}
This is typical for exciton hopping between domains in molecular aggregates polymers \citep{scholes2006excitons}. 

\subsubsection{Charge separation the photosystems must also be fast}

The reaction centres, $PS_m$, will only function effectively is charge separation competes with exciton hopping,
\begin{align}
    \frac{1}{k_\text{trap}}\sim \tau_{hop}
\end{align}
This ensures that charge separation can occur before the exciton can escape back into the antenna. In Earth's oxygenic organisms, $\sim 10$ ps is approximately the timescale of primary charge separation in PSII \citep{duan2017primary}.

\subsubsection{Electron transfer between photosystems is likely the rate limiting step}

The transfer of electrons between different photosystems $PS_m$ and $PS_{m+1}$ likely involves the diffusion of an electron carrier across a relatively long distance. We will assume,
\begin{align}
 \frac{1}{k_\text{lin}}\sim 10 \text{ ms}
\end{align}
In the case our oxygenic model, this means that $PS_{ox}$ process a maximum of 100 photons per second. By the same logic we also assume,
\begin{align}
    \frac{1}{k_\text{cyc}}=\frac{1}{k_\text{red}}\sim 10 \text{ ms}
\end{align}

\subsubsection{Donor oxidation is faster than linear electron flow}

$PS_{ox}$ is the oxidizing photosystem, meaning that the pigment $P_{ox}$ must have a very high ionization potential (extremely so for oxygenic photosynthesis). In other words, the species $P_{ox}^+$ \emph{must} be highly oxidizing. Such an strong oxidizing agent poses dangers to any biological system, meaning that there will likely be a strong selection pressure to ensure that $P_{ox}^+$ does not accumulate. This is most easily achieved by ensuring that donor oxidation is faster than linear electron flow. We assume that,
\begin{align}
k_\text{trap} >> k_\text{ox} > k_\text{lin}
\end{align}
and pick the arbitrary value of,
\begin{align}
    \frac{1}{k_{ox}}\sim 1 \text{ ms}
\end{align}
Though very crude, this seems to align with real oxygenic photosynthesis: The various one-photon steps in the Kok (S-state) catalytic cycle in PSII occur on timescales of $100 - 1400$ $\mu$s\citet{rezarazeghifard1997}. $P680^+$ in PSII is indeed an exceptionally strong oxidizing agent and its accumulation due to intrerupted water oxidation is a source of photodamage. 

\subsubsection{Non-radiative charge recombination is extremely slow}

If charge recombination in the photosystems were fast, or even comparable to the other redox processes, the system would simply not function. We therefore assume that, as on Earth, there would be an extremely strong selection pressure to ensure that charge recombination was extremely slow. In real systems it occurs on a timescale of,
\begin{align}
\frac{1}{k_\text{recom}}\sim 100 \text{ ms} -1 \text{ s}    
\end{align}
which is two orders of magnitude slower than the next slowest process. We could therefore assume that,
\begin{align}
\frac{1}{k_\text{recom}}\rightarrow \infty    
\end{align}
unless we are specifically studying the kinetics of recombination, possibly as a source of photodamage in very high light. 

\section{Photosynthetic properties generated by the generate algorithm}

For a given photosynthetic model $\mathcal{M}$ in a set of environmental conditions defined by $E_{e,\lambda}(a,T_s,R_s)$, the genetic algorithm yields a population of 500 'evolved' parameterizations (genomes) of the light-harvesting antenna model. Averaging over this population gives a number of photosynthetic properties which are listed in the following tables. 

\begin{table*}[ht!]
\caption {A summary of the photosynthetic properties output by the genetic algorithm for an exoplanet orbiting an M-type star ($T_s=2343$ K, $M_s = 0.08$ $M_\odot$). $\langle \nu\rangle$ is the $CO_2$ reduction rate, averaged over the final evolved population, and $\sigma_\nu$ is the standard deviation. Similarly, $\langle f\rangle $ and $\sigma_f$ are the average and standard deviation of the fitness parameter. Finally, $\langle N_b^a \sum N_{b,p}^a \rangle$ and $\sigma_{N_b^a \sum N_{b,p}^a}$ are the average and standard deviation of the antenna \emph{size}, where $N_b^a$ is the numbers of \emph{branches} in the antenna and $N_{b,p}^a$ is the number of pigments per branch.}
\label{table:1} 
\centering
\begin{tabular}{cccccccccc}
\hline\hline
$\mathcal{M}$ & $T_s $ (K) & $M_s$ $(M_\odot)$ &	$a$ (AU) & $\left< \nu \right>$ ($s^{-1}$)	 &$\sigma_{\nu}$ ($s^{-1}$) &$\left< f \right>$	($s^{-1}$)& $\sigma_{f}$ ($s^{-1}$)& $ \left< N_b^a \sum N_{b,p}^a \right> $ &	$ \sigma_{N_b^a \sum N_{b,p}^a} $\\
 \hline
 Anoxygenic& 2343& 0.08& 0.017& 53.7& 8.1& 29.3& 5.5& 1584.3& 495.6\\ 
Anoxygenic& 2343& 0.08& 0.018& 55.2& 7.9& 29.3& 6.0& 1789.1& 406.7\\ 
Anoxygenic& 2343& 0.08& 0.018& 52.2& 7.9& 27.2& 6.0& 1769.5& 368.0\\ 
Anoxygenic& 2343& 0.08& 0.019& 48.2& 9.3& 26.0& 6.2& 1535.7& 446.8\\ 
Anoxygenic& 2343& 0.08& 0.020& 47.2& 7.7& 24.7& 5.4& 1603.1& 416.3\\ 
Anoxygenic& 2343& 0.08& 0.021& 48.9& 7.7& 24.4& 4.9& 1849.3& 480.8\\ 
Anoxygenic& 2343& 0.08& 0.021& 44.6& 7.0& 23.5& 4.5& 1558.9& 361.3\\ 
Anoxygenic& 2343& 0.08& 0.022& 41.1& 6.7& 20.9& 4.6& 1500.9& 403.1\\ 
Anoxygenic& 2343& 0.08& 0.023& 39.4& 7.6& 20.1& 4.5& 1461.2& 391.8\\ 
Anoxygenic& 2343& 0.08& 0.024& 40.4& 7.5& 19.7& 3.8& 1631.1& 417.1\\ 
Anoxygenic& 2343& 0.08& 0.025& 36.6& 8.2& 16.7& 4.5& 1586.0& 426.4\\ 
Anoxygenic& 2343& 0.08& 0.026& 34.8& 6.7& 16.0& 3.4& 1504.1& 435.2\\ 
Anoxygenic& 2343& 0.08& 0.027& 38.6& 5.8& 15.9& 3.6& 1918.8& 326.0\\ 
Anoxygenic& 2343& 0.08& 0.028& 36.9& 6.2& 13.7& 3.3& 1997.3& 437.2\\ 
Anoxygenic& 2343& 0.08& 0.029& 30.7& 5.2& 12.6& 2.6& 1510.7& 309.4\\ 
Anoxygenic& 2343& 0.08& 0.030& 28.8& 4.3& 11.7& 2.2& 1426.5& 229.3\\ 
Anoxygenic& 2343& 0.08& 0.031& 28.9& 6.2& 10.2& 2.3& 1619.2& 465.8\\ 
Anoxygenic& 2343& 0.08& 0.032& 26.3& 5.8& 8.8& 2.3& 1525.7& 405.5\\ 
Anoxygenic& 2343& 0.08& 0.034& 25.3& 5.5& 7.6& 1.8& 1547.4& 408.3\\ 
Anoxygenic& 2343& 0.08& 0.035& 22.1& 4.4& 6.5& 1.4& 1350.7& 348.7\\ 
\hline
Oxygenic& 2343& 0.08& 0.017& 23.7& 4.1& 10.6& 2.4& 854.1& 215.0\\ 
Oxygenic& 2343& 0.08& 0.018& 24.5& 5.5& 9.9& 2.6& 1042.3& 348.9\\ 
Oxygenic& 2343& 0.08& 0.018& 22.1& 3.9& 9.3& 2.0& 893.3& 205.6\\ 
Oxygenic& 2343& 0.08& 0.019& 22.2& 3.8& 8.8& 1.9& 990.3& 218.7\\ 
Oxygenic& 2343& 0.08& 0.020& 20.3& 3.5& 7.4& 1.8& 955.9& 229.8\\ 
Oxygenic& 2343& 0.08& 0.021& 19.2& 4.1& 6.7& 1.5& 941.2& 309.2\\ 
Oxygenic& 2343& 0.08& 0.021& 16.9& 3.1& 5.7& 1.1& 825.6& 202.1\\ 
Oxygenic& 2343& 0.08& 0.022& 18.1& 2.7& 5.1& 1.2& 1032.4& 188.0\\ 
Oxygenic& 2343& 0.08& 0.023& 15.1& 2.1& 4.3& 0.8& 827.6& 134.2\\ 
Oxygenic& 2343& 0.08& 0.024& 12.8& 2.3& 3.4& 0.9& 714.5& 143.8\\ 
Oxygenic& 2343& 0.08& 0.025& 11.5& 2.4& 2.9& 0.7& 654.0& 176.1\\ 
Oxygenic& 2343& 0.08& 0.026& 10.1& 1.5& 2.2& 0.5& 590.2& 92.2\\ 
Oxygenic& 2343& 0.08& 0.027& 8.7& 1.3& 1.8& 0.4& 506.1& 77.9\\ 
Oxygenic& 2343& 0.08& 0.028& 7.6& 1.8& 1.4& 0.3& 449.6& 155.7\\ 
Oxygenic& 2343& 0.08& 0.029& 7.0& 1.3& 1.0& 0.2& 438.4& 104.0\\ 
Oxygenic& 2343& 0.08& 0.030& 4.5& 0.6& 0.8& 0.2& 219.7& 50.8\\ 
Oxygenic& 2343& 0.08& 0.031& 4.0& 0.6& 0.6& 0.1& 198.1& 54.8\\ 
Oxygenic& 2343& 0.08& 0.032& 3.0& 0.3& 0.5& 0.1& 117.7& 30.2\\ 
Oxygenic& 2343& 0.08& 0.034& 2.5& 0.3& 0.4& 0.1& 90.9& 27.5\\ 
Oxygenic& 2343& 0.08& 0.035& 2.4& 0.3& 0.3& 0.1& 98.3& 30.1\\ 
\hline
Hypothetical& 2343& 0.08& 0.017& 40.9& 5.2& 18.6& 3.4& 771.9& 250.7\\ 
Hypothetical& 2343& 0.08& 0.018& 42.0& 4.9& 18.1& 3.7& 1015.2& 280.1\\ 
Hypothetical& 2343& 0.08& 0.018& 39.8& 5.4& 17.7& 3.4& 922.1& 378.7\\ 
Hypothetical& 2343& 0.08& 0.019& 38.5& 4.6& 17.8& 3.3& 855.4& 251.8\\ 
Hypothetical& 2343& 0.08& 0.020& 38.1& 5.1& 18.2& 3.0& 853.5& 279.5\\ 
Hypothetical& 2343& 0.08& 0.021& 37.8& 5.8& 17.0& 3.5& 1014.0& 384.2\\ 
Hypothetical& 2343& 0.08& 0.022& 36.0& 4.7& 17.2& 3.0& 901.8& 244.0\\ 
Hypothetical& 2343& 0.08& 0.023& 34.7& 5.2& 16.4& 3.3& 985.4& 363.2\\ 
Hypothetical& 2343& 0.08& 0.025& 32.0& 4.7& 15.0& 3.1& 967.9& 283.1\\ 
Hypothetical& 2343& 0.08& 0.028& 30.9& 4.2& 13.2& 2.6& 1138.1& 248.6\\ 
Hypothetical& 2343& 0.08& 0.030& 30.1& 4.2& 12.2& 2.5& 1247.5& 234.4\\ 
Hypothetical& 2343& 0.08& 0.032& 25.0& 4.2& 10.1& 2.2& 1022.7& 294.2\\ 
Hypothetical& 2343& 0.08& 0.035& 22.3& 3.7& 8.6& 1.7& 969.5& 235.7\\ 
\hline
\end{tabular}
\end{table*}

\begin{table*}[ht!]
\caption {The same as Table \ref{table:1} but for a K-type star ($T_s=3416$ K, $M_s = 0.3$ $M_\odot$).}
\label{table:2} 
\centering
\begin{tabular}{cccccccccc}
\hline\hline
$\mathcal{M}$ & $T_s $ (K) & $M_s$ $(M_\odot)$ &	$a$ (AU) & $\left< \nu \right>$ ($s^{-1}$)	 &$\sigma_{\nu}$ ($s^{-1}$) &$\left< f \right>$	($s^{-1}$)& $\sigma_{f}$ ($s^{-1}$)& $ \left< N_b^a \sum N_{b,p}^a \right> $ &	$ \sigma_{N_b^a \sum N_{b,p}^a} $\\
 \hline
Anoxygenic& 3416& 0.3& 0.107& 64.62& 8.40& 37.10& 6.50& 1139.86& 315.96\\ 
Anoxygenic& 3416& 0.3& 0.111& 66.26& 8.99& 37.44& 6.80& 1361.13& 433.18\\ 
Anoxygenic& 3416& 0.3& 0.115& 65.48& 8.02& 37.35& 6.85& 1377.01& 357.80\\ 
Anoxygenic& 3416& 0.3& 0.119& 64.34& 8.44& 37.44& 6.01& 1346.99& 378.74\\ 
Anoxygenic& 3416& 0.3& 0.124& 63.20& 9.52& 35.36& 7.33& 1520.74& 471.27\\ 
Anoxygenic& 3416& 0.3& 0.128& 59.80& 8.01& 35.55& 5.90& 1237.09& 367.25\\ 
Anoxygenic& 3416& 0.3& 0.133& 59.87& 9.17& 33.97& 7.49& 1477.76& 338.68\\ 
Anoxygenic& 3416& 0.3& 0.137& 59.93& 9.10& 34.07& 6.54& 1544.11& 411.64\\ 
Anoxygenic& 3416& 0.3& 0.142& 60.04& 9.39& 33.68& 6.46& 1657.99& 537.68\\ 
Anoxygenic& 3416& 0.3& 0.147& 57.03& 8.13& 32.94& 5.81& 1491.51& 342.86\\ 
Anoxygenic& 3416& 0.3& 0.153& 54.08& 8.54& 29.81& 6.90& 1566.70& 391.67\\ 
Anoxygenic& 3416& 0.3& 0.158& 52.68& 9.08& 28.90& 6.60& 1577.52& 503.43\\ 
Anoxygenic& 3416& 0.3& 0.164& 50.04& 7.79& 28.18& 6.64& 1424.37& 298.95\\ 
Anoxygenic& 3416& 0.3& 0.170& 50.83& 7.36& 28.06& 5.60& 1571.06& 310.04\\ 
Anoxygenic& 3416& 0.3& 0.176& 51.37& 8.27& 27.19& 5.33& 1753.93& 486.06\\ 
Anoxygenic& 3416& 0.3& 0.182& 45.22& 7.85& 24.80& 4.95& 1421.29& 432.90\\ 
Anoxygenic& 3416& 0.3& 0.188& 48.10& 7.50& 25.31& 4.78& 1698.13& 418.69\\ 
Anoxygenic& 3416& 0.3& 0.195& 46.31& 6.29& 24.62& 4.27& 1630.49& 277.00\\ 
Anoxygenic& 3416& 0.3& 0.202& 46.13& 6.94& 23.28& 4.83& 1778.72& 413.93\\ 
Anoxygenic& 3416& 0.3& 0.209& 42.31& 6.60& 21.09& 4.59& 1651.67& 302.76\\ 
\hline
Oxygenic& 3416& 0.3& 0.107& 43.36& 6.21& 20.18& 4.34& 830.72& 273.86\\ 
Oxygenic& 3416& 0.3& 0.111& 44.26& 5.54& 21.49& 4.02& 870.60& 254.17\\ 
Oxygenic& 3416& 0.3& 0.115& 43.76& 5.83& 21.86& 4.08& 870.86& 298.83\\ 
Oxygenic& 3416& 0.3& 0.119& 44.44& 6.06& 21.36& 4.35& 1067.06& 252.51\\ 
Oxygenic& 3416& 0.3& 0.124& 42.71& 6.20& 21.13& 4.24& 988.39& 278.25\\ 
Oxygenic& 3416& 0.3& 0.128& 42.91& 5.60& 21.48& 3.52& 1038.03& 352.76\\ 
Oxygenic& 3416& 0.3& 0.133& 40.96& 5.80& 20.10& 4.40& 1047.45& 245.08\\ 
Oxygenic& 3416& 0.3& 0.137& 42.13& 5.72& 19.42& 4.38& 1300.01& 359.68\\ 
Oxygenic& 3416& 0.3& 0.142& 38.32& 5.57& 18.99& 3.57& 1021.07& 285.16\\ 
Oxygenic& 3416& 0.3& 0.147& 38.06& 6.00& 18.60& 4.18& 1088.28& 284.64\\ 
Oxygenic& 3416& 0.3& 0.153& 37.79& 5.20& 18.27& 3.55& 1144.16& 323.19\\ 
Oxygenic& 3416& 0.3& 0.158& 32.61& 5.75& 16.57& 3.69& 852.41& 248.28\\ 
Oxygenic& 3416& 0.3& 0.164& 35.41& 5.29& 16.25& 3.87& 1212.99& 259.74\\ 
Oxygenic& 3416& 0.3& 0.170& 34.77& 5.43& 16.97& 3.23& 1119.68& 289.02\\ 
Oxygenic& 3416& 0.3& 0.176& 33.89& 5.42& 15.86& 3.34& 1187.83& 332.20\\ 
Oxygenic& 3416& 0.3& 0.182& 29.14& 5.53& 14.15& 3.35& 917.07& 264.95\\ 
Oxygenic& 3416& 0.3& 0.188& 30.73& 5.24& 14.29& 2.85& 1108.09& 304.21\\ 
Oxygenic& 3416& 0.3& 0.195& 29.13& 4.66& 12.95& 2.85& 1112.59& 269.51\\ 
Oxygenic& 3416& 0.3& 0.202& 30.05& 4.85& 12.88& 2.55& 1248.79& 290.90\\ 
Oxygenic& 3416& 0.3& 0.209& 27.90& 5.31& 11.79& 2.71& 1180.33& 318.60\\ 
\hline
Hypothetical& 3416& 0.3& 0.107& 37.90& 5.66& 15.95& 3.71& 352.26& 211.90\\ 
Hypothetical& 3416& 0.3& 0.116& 43.93& 4.67& 18.36& 3.33& 915.43& 226.04\\ 
Hypothetical& 3416& 0.3& 0.125& 40.47& 4.90& 18.60& 3.56& 727.39& 195.17\\ 
Hypothetical& 3416& 0.3& 0.134& 38.41& 4.27& 18.25& 3.13& 718.89& 170.03\\ 
Hypothetical& 3416& 0.3& 0.145& 37.30& 5.26& 17.79& 3.38& 808.32& 262.63\\ 
Hypothetical& 3416& 0.3& 0.156& 37.53& 4.58& 17.80& 2.68& 966.10& 271.60\\ 
Hypothetical& 3416& 0.3& 0.168& 38.79& 4.30& 16.77& 2.94& 1319.54& 312.80\\ 
Hypothetical& 3416& 0.3& 0.181& 34.82& 5.08& 16.36& 2.93& 1085.94& 286.63\\ 
Hypothetical& 3416& 0.3& 0.194& 31.79& 4.64& 14.72& 2.89& 1042.37& 242.59\\ 
Hypothetical& 3416& 0.3& 0.209& 28.02& 4.78& 12.74& 2.71& 947.28& 247.39\\ 
\hline
\end{tabular}
\end{table*}

\begin{table*}[ht!]
\caption {The same as Table \ref{table:1} but for a G-type star ($T_s=5697$ K, $M_s = 1.0$ $M_\odot$).}
\label{table:3} 
\centering
\begin{tabular}{cccccccccc}
\hline\hline
$\mathcal{M}$ & $T_s $ (K) & $M_s$ $(M_\odot)$ &	$a$ (AU) & $\left< \nu \right>$ ($s^{-1}$)	 &$\sigma_{\nu}$ ($s^{-1}$) &$\left< f \right>$	($s^{-1}$)& $\sigma_{f}$ ($s^{-1}$)& $ \left< N_b^a \sum N_{b,p}^a \right> $ &	$ \sigma_{N_b^a \sum N_{b,p}^a} $\\
 \hline
Anoxygenic& 5697& 1.0& 0.923& 72.0& 8.0& 39.8& 6.7& 1487.7& 380.2\\ 
Anoxygenic& 5697& 1.0& 0.951& 65.7& 10.2& 36.1& 9.0& 1321.9& 355.2\\ 
Anoxygenic& 5697& 1.0& 0.980& 69.8& 8.5& 39.7& 6.4& 1454.4& 363.1\\ 
Anoxygenic& 5697& 1.0& 1.010& 68.1& 10.1& 37.8& 8.5& 1546.0& 396.7\\ 
Anoxygenic& 5697& 1.0& 1.040& 66.0& 9.8& 38.0& 7.7& 1388.1& 412.4\\ 
Anoxygenic& 5697& 1.0& 1.072& 67.0& 11.0& 37.6& 8.2& 1620.1& 485.6\\ 
Anoxygenic& 5697& 1.0& 1.105& 62.9& 8.4& 37.2& 6.4& 1299.2& 349.9\\ 
Anoxygenic& 5697& 1.0& 1.138& 61.4& 9.0& 36.0& 6.7& 1320.5& 419.7\\ 
Anoxygenic& 5697& 1.0& 1.173& 62.4& 8.9& 36.6& 6.9& 1448.0& 464.2\\ 
Anoxygenic& 5697& 1.0& 1.209& 61.7& 9.4& 35.4& 7.1& 1546.3& 442.9\\ 
Anoxygenic& 5697& 1.0& 1.245& 61.3& 11.6& 34.5& 7.1& 1659.9& 644.2\\ 
Anoxygenic& 5697& 1.0& 1.283& 61.5& 8.1& 35.7& 5.7& 1608.4& 479.7\\ 
Anoxygenic& 5697& 1.0& 1.322& 57.9& 9.7& 32.9& 7.4& 1586.5& 431.1\\ 
Anoxygenic& 5697& 1.0& 1.362& 56.2& 10.3& 32.6& 7.1& 1492.5& 462.0\\ 
Anoxygenic& 5697& 1.0& 1.404& 58.7& 10.2& 31.4& 6.9& 1907.8& 647.1\\ 
Anoxygenic& 5697& 1.0& 1.447& 55.2& 8.0& 30.6& 6.6& 1680.0& 384.6\\ 
Anoxygenic& 5697& 1.0& 1.491& 51.3& 8.3& 30.6& 5.6& 1341.6& 319.5\\ 
Anoxygenic& 5697& 1.0& 1.536& 52.0& 9.7& 29.1& 6.0& 1599.2& 598.2\\ 
Anoxygenic& 5697& 1.0& 1.583& 50.6& 8.4& 28.7& 6.0& 1539.2& 366.2\\ 
Anoxygenic& 5697& 1.0& 1.631& 55.3& 8.2& 28.1& 6.5& 2102.5& 429.1\\ 
\hline
Oxygenic& 5697& 1.0& 0.923& 50.8& 5.9& 20.1& 4.4& 762.2& 287.3\\ 
Oxygenic& 5697& 1.0& 0.951& 49.2& 5.8& 21.0& 4.0& 609.4& 213.7\\ 
Oxygenic& 5697& 1.0& 0.980& 51.4& 5.3& 22.0& 3.9& 812.5& 297.1\\ 
Oxygenic& 5697& 1.0& 1.010& 49.7& 6.1& 21.8& 3.9& 756.8& 288.4\\ 
Oxygenic& 5697& 1.0& 1.040& 49.2& 5.8& 21.9& 4.4& 785.0& 209.1\\ 
Oxygenic& 5697& 1.0& 1.072& 48.2& 7.4& 21.8& 4.8& 796.7& 336.7\\ 
Oxygenic& 5697& 1.0& 1.105& 48.1& 5.7& 22.1& 4.0& 823.4& 321.5\\ 
Oxygenic& 5697& 1.0& 1.138& 49.1& 6.6& 22.5& 4.8& 974.9& 333.9\\ 
Oxygenic& 5697& 1.0& 1.173& 46.1& 5.8& 21.5& 4.3& 845.8& 281.8\\ 
Oxygenic& 5697& 1.0& 1.209& 49.0& 5.9& 22.7& 4.1& 1094.2& 366.2\\ 
Oxygenic& 5697& 1.0& 1.245& 47.0& 6.2& 22.1& 4.8& 1022.0& 391.0\\ 
Oxygenic& 5697& 1.0& 1.283& 47.3& 5.6& 22.7& 4.2& 1069.1& 303.2\\ 
Oxygenic& 5697& 1.0& 1.322& 44.5& 5.9& 22.2& 4.6& 902.9& 298.9\\ 
Oxygenic& 5697& 1.0& 1.362& 44.0& 6.9& 22.1& 4.7& 932.2& 281.0\\ 
Oxygenic& 5697& 1.0& 1.404& 44.3& 6.6& 22.2& 4.2& 1020.1& 331.3\\ 
Oxygenic& 5697& 1.0& 1.447& 40.9& 5.9& 20.4& 4.7& 909.6& 315.2\\ 
Oxygenic& 5697& 1.0& 1.491& 42.1& 6.9& 21.3& 4.6& 1003.8& 295.5\\ 
Oxygenic& 5697& 1.0& 1.536& 41.0& 6.7& 20.5& 4.9& 1027.1& 295.3\\ 
Oxygenic& 5697& 1.0& 1.583& 41.6& 6.8& 20.9& 4.1& 1108.3& 442.8\\ 
Oxygenic& 5697& 1.0& 1.631& 38.9& 6.1& 19.6& 4.3& 1013.9& 351.9\\ 
\hline
Hypothetical& 5697& 1.0& 0.923& 41.9& 4.5& 19.4& 3.0& 742.3& 207.8\\ 
Hypothetical& 5697& 1.0& 0.983& 41.4& 5.7& 19.0& 3.5& 887.4& 310.5\\ 
Hypothetical& 5697& 1.0& 1.047& 37.3& 6.4& 17.4& 4.2& 781.0& 328.5\\ 
Hypothetical& 5697& 1.0& 1.116& 38.3& 5.4& 17.5& 3.9& 983.2& 317.2\\ 
Hypothetical& 5697& 1.0& 1.189& 35.7& 5.4& 17.1& 3.4& 886.9& 254.4\\ 
Hypothetical& 5697& 1.0& 1.266& 37.6& 5.5& 16.3& 3.3& 1251.8& 339.8\\ 
Hypothetical& 5697& 1.0& 1.349& 35.2& 6.1& 15.9& 3.7& 1158.1& 354.6\\ 
Hypothetical& 5697& 1.0& 1.437& 33.4& 4.6& 15.1& 3.0& 1130.8& 317.7\\ 
Hypothetical& 5697& 1.0& 1.531& 32.6& 4.6& 14.2& 2.8& 1216.2& 257.8\\ 
Hypothetical& 5697& 1.0& 1.631& 30.9& 4.5& 13.0& 2.5& 1241.2& 288.5\\ 
\hline
\end{tabular}
\end{table*}

\end{appendix}

\bsp	
\label{lastpage}

\end{document}